\documentclass[12pt]{article}
\RequirePackage[OT1]{fontenc}
\usepackage{amsthm,amsmath,amssymb,amsfonts,bm,enumerate,xcolor,graphicx,natbib,color,url}
\RequirePackage[colorlinks,citecolor=blue,urlcolor=blue]{hyperref}
\usepackage{hyperref}
\usepackage{subcaption} 
\usepackage{ulem}
\usepackage{pifont}
\usepackage{textcmds}
\usepackage{booktabs}
\usepackage{multirow}
\usepackage{booktabs}
\usepackage{adjustbox}
\usepackage[usestackEOL]{stackengine}
\usepackage[toc,page]{appendix} 
\usepackage{algorithm,algorithmic}
\usepackage{lipsum}
\usepackage{setspace}
\usepackage{subcaption,dcolumn}
\usepackage{array}
\usepackage{longtable}
\usepackage{multirow}

\usepackage{setspace}
\allowdisplaybreaks

\definecolor{lb}{RGB}{44,139,183}

\usepackage{ulem} 

\def\frac#1#2{{\textstyle{#1\over#2}}}

\DeclareSymbolFont{AMSb}{U}{msb}{m}{n}
\DeclareMathSymbol{\Natural}{\mathbin}{AMSb}{"4E}
\DeclareMathSymbol{\Integer}{\mathbin}{AMSb}{"5A}
\DeclareMathSymbol{\Real}{\mathbin}{AMSb}{"52}
\DeclareMathSymbol{\Rational}{\mathbin}{AMSb}{"51}
\DeclareMathSymbol{\Imaginary}{\mathbin}{AMSb}{"49}
\DeclareMathSymbol{\Complex}{\mathbin}{AMSb}{"43} 
\DeclareMathSymbol{\Disk}{\mathbin}{AMSb}{"44} 
\def\bi{\begin{itemize}}
\def\ei{\end{itemize}}
\def\bd{\begin{description}}
\def\ed{\end{description}}
\def\ben{\begin{enumerate}}
\def\een{\end{enumerate}}
\def\bv{\begin{verbatim}}
\def\ev{\end{verbatim}}

\def\hash{{\#}}

\def\hat#1{{\widehat{#1}}}

\def\2to{{\ {\buildrel 2\over \longrightarrow}\ }}

\def\I1ton{{$I_1,\ldots,I_n$}}
\def\X1ton{{$X_1,\ldots,X_n$}}
\def\Y1ton{{$Y_1,\ldots,Y_n$}}
\def\Z1ton{{$Z_1,\ldots,Z_n$}}
\def\R1ton{{$R_1,\ldots,R_n$}}
\def\e1ton{{$e_1,\ldots,e_n$}}
\def\t1ton{{$t_1,\ldots,t_n$}}
\def\x1ton{{$x_1,\ldots,x_n$}}
\def\y1ton{{$y_1,\ldots,y_n$}}
\def\z1ton{{$z_1,\ldots,z_n$}}

\def\tt#1{{\texttt{#1}}}

\def\tt#1{{\texttt{#1}}}

\newcommand{\ffrac}[2]{\ensuremath{\frac{\displaystyle #1}{\displaystyle #2}}}

\newcommand{\blind}{1}

\colorlet{shadecolor}{gray!40}

\begin{document}
\thispagestyle{empty}
\baselineskip=28pt
\vskip 5mm

\begin{center} {\Large{\bf {\color{black}{Exploring the Efficacy of Statistical and Deep Learning Methods for Large Spatial Datasets: A Case Study}} }}
\end{center}


\baselineskip=12pt

\vskip 5mm

\if1\blind
{
\begin{center}
\large
Arnab Hazra$^1$, Pratik Nag$^2$, Rishikesh Yadav$^3$,  Ying Sun$^2$\\ 
\end{center}
\footnotetext[1]{
\baselineskip=10pt Department of Mathematics and Statistics, Indian Institute of Technology Kanpur, Kanpur 208016, India. Email: ahazra@iitk.ac.in}
\footnotetext[2]{
\baselineskip=10pt Statistics Program, Computer, Electrical and Mathematical Sciences and Engineering (CEMSE) Division, King Abdullah University of Science and Technology (KAUST), Thuwal, Saudi Arabia. E-mails: pratik.nag@kaust.edu.sa; Ying.sun@kaust.edu.sa}
\footnotetext[3]{Department of Decision Science, HEC Montreal, Montreal, Canada. Email: rishikesh.yadav@hec.ca }
} \fi


\baselineskip=26pt
\vskip 2mm



\begin{center}
    {\large{\bf Abstract}}
\end{center}
\baselineskip=26pt



\noindent  Increasingly large and complex spatial datasets pose massive inferential challenges due to high computational and storage costs. Our study is motivated by the KAUST Competition on Large Spatial Datasets 2023, which tasked participants with estimating spatial covariance-related parameters and predicting values at testing sites, along with uncertainty estimates. We compared various statistical and deep learning approaches through cross-validation and ultimately selected the Vecchia approximation technique for model fitting. To overcome the constraints in the \texttt{R} package \texttt{GpGp}, which lacked support for fitting zero-mean Gaussian processes and direct uncertainty estimation--two things that are necessary for the competition, we developed additional \texttt{R} functions. Besides, we implemented certain subsampling-based approximations and parametric smoothing for skewed sampling distributions of the estimators. Our team \textit{DesiBoys} secured victory in two out of four sub-competitions, validating the effectiveness of our proposed strategies. Moreover, we extended our evaluation to a large real spatial satellite-derived dataset on total precipitable water, where we compared the predictive performances of different models using multiple diagnostics.




\baselineskip=26pt
\section{Introduction} 
\label{sec:introduction}

Gaussian processes (GPs) are the most popular model in spatial geostatistics due to their several attractive theoretical and computational advantages \citep{gelfand2016spatial}. A spatial GP, say, $Y(\bm{s}), \bm{s} \in \mathcal{D} \subset \mathbb{R}^2$ is characterized by a mean function $\mu(\cdot)$ and a covariance kernel $K(\cdot, \cdot)$. For the set of observation locations $\mathcal{S} = \{ \bm{s}_1, \ldots, \bm{s}_N \}$ and the corresponding observations $\{Y(\bm{s}_1), \ldots, Y(\bm{s}_N)\}$, the random vector $\bm{Y} = [Y(\bm{s}_1), \ldots, Y(\bm{s}_N)]'$ follows an $N$-variate multivariate Gaussian distribution with its density given by
\begin{equation} \label{eq:multivariate_normal}
    f_{\bm{Y}}(\bm{y}) = \ffrac{1}{\sqrt{\vert 2 \pi \bm{\Sigma} \vert }} \exp \left[- \ffrac{1}{2} (\bm{y} - \bm{\mu})' \bm{\Sigma}^{-1} (\bm{y} - \bm{\mu}) \right], \quad \bm{y} \in \mathbb{R}^N,
\end{equation}
where $\bm{\mu} = [\mu(\bm{s}_1), \ldots, \mu(\bm{s}_N)]'$ is the mean vector and $\bm{\Sigma}$ is the $(N \times N)$-dimensional covariance matrix with its $(i,j)^{th}$ entry given by $\Sigma_{ij} = K(\bm{s}_i, \bm{s}_j)$ for $i,j=1, \ldots, N$. Thus, the inferences depend only on the first two moments of the data. Specifically, here we need to calculate the determinant and the inverse of $\bm{\Sigma}$ at every step of a computing approach, e.g., the Newton-Raphson algorithm, and it requires $\mathcal{O}(N^3)$ many operations {\color{black}{with storage cost  $\mathcal{O}(N^2)$}}. Several linear algebraic techniques are available for the mean vector and the covariance matrix-related computations and approximations when $N$ is moderately large; see \cite{press2007numerical} and \cite{ trefethen2022numerical}.

Inferences based on unstructured $\bm{\mu}$ and $\bm{\Sigma}$ in \eqref{eq:multivariate_normal} involve difficulties due to the number of parameters being larger than the data dimension leading to clear identifiability issues. Among different parameterizations of $\bm{\mu}$ and $\bm{\Sigma}$, while imposing different spatial regression structures on $\bm{\mu}$ is easier, finding parameterizations that allow $\bm{\Sigma}$ to be positive definite is non-trivial. A popular choice for the covariance kernel $K(\cdot, \cdot)$ is the Mat\'ern correlation function \citep{genton2001classes, williams2006gaussian} given by
\begin{equation} \label{eq:matern_correlation}
\textrm{Cov}[Y(\bm{s}), Y(\bm{s}')] = \sigma^2 \ffrac{2^{1-\nu}}{\Gamma(\nu)} \left( \ffrac{\Vert \bm{s} - \bm{s}' \Vert}{\phi} \right)^{\nu} K_{\nu} \left( \ffrac{\Vert \bm{s} - \bm{s}' \Vert}{\phi} \right) + \tau^2 \mathbb{I}(\bm{s} = \bm{s}'),
\end{equation}
where $\Vert \bm{s} - \bm{s}' \Vert$ is the Euclidean distance between $\bm{s}$ and $\bm{s}'$, and the positive-valued parameters $\sigma^2$, $\phi$, $\nu$, and $\tau^2$ are partial sill, range, smoothness, and nugget, respectively. In \eqref{eq:matern_correlation}, $K_{\nu}(\cdot)$ is the modified Bessel function of second kind with degree $\nu$, and $\mathbb{I}(\bm{s} = \bm{s}') = 1$ if $\bm{s} = \bm{s}'$, and 0 otherwise. Except for some trivial choices of the data locations and the Mat\'ern parameter values, the matrix $\bm{\Sigma}$ constructed following \eqref{eq:matern_correlation} and its inverse are dense.


While computing \eqref{eq:multivariate_normal} with covariance structure \eqref{eq:matern_correlation} is feasible for moderate $N \approx 10^3,10^4$, the large irregularly spaced spatial datasets with $N \approx 10^5,10^6$, or even more, obtained from various scientific experiments and satellites in modern times, still pose a challenge in utilizing GPs due to their excessive computational complexity and storage. Early attempts to address this problem involved employing pseudo-likelihoods \citep{besag1977efficiency}, factoring the density into conditional distributions \citep{vecchia1988estimation}, utilizing spectral domain modeling \citep{stein1995fixed}, or implementing tapered covariance functions \citep{gaspari1999construction}. In the late 2000s, alternative methods based on low-rank approximations of GPs gained popularity. These included discrete process convolutions, fixed rank kriging, predictive processes, lattice kriging, and stochastic partial differential equations. Extensive reviews by \cite{sun2012geostatistics}, \cite{bradley2016comparison}, \cite{liu2020gaussian}, and \cite{cressie2022basis} demonstrated the effectiveness of these approaches in modeling spatial data. However, over time, scientists noticed limitations in many of these approximation methods, such as their tendency to oversmooth the data \citep{simpson2012order} and constraints on the size of the datasets they could handle. Besides, some of these approaches (fixed rank kriging, for example) can perform kriging only and cannot provide inferences about parameters involved in parameterized forms of $\bm{\Sigma}$, e.g., the Mat\'ern correlation parameters in \eqref{eq:matern_correlation}.



As a result, recent research in this field has focused on efficiently utilizing modern computing platforms and developing parallelizable methods. For instance, 
\cite{eidsvik2014estimation} proposed a block composite likelihood approach where the composite likelihood \citep{lindsay1988composite} is constructed from the joint densities of pairs of adjacent spatial blocks. \cite{paciorek2013parallelizing} explored parallel computing techniques to calculate the likelihood in \eqref{eq:multivariate_normal} more efficiently. 
\cite{katzfuss2017multi} and \cite{katzfuss2017parallel} introduced a basis-function approach suitable for distributed computing. Alternatively, \cite{barbian2017spatial,guhaniyogi2018meta} proposed dividing the data into numerous subsets, conducting parallel inference on each subset, and then combining the results. \textcolor{black}{\cite{sun2016statistically} 
proposed new unbiased estimating equations based on score equation approximations that are both computationally and statistically efficient. \cite{stein2004approximating} and \cite{datta2016nonseparable} demonstrated the adaptability of \cite{vecchia1988estimation} to approximate the restricted likelihood and assessed the efficacy of the resulting approximation using an estimating equations approach, considering conditioning on the observations closest to the one whose conditional density is being approximated.} 

In recent years, the widespread popularity of deep neural networks (DNNs) in environmental applications has led to their increasing adoption in spatial problems due to their distribution-free characteristics and scalability to large datasets \citep{najafabadi2015deep, wikle2023statistical}. Researchers have explored various approaches to incorporate DNNs in spatial applications; for instance, \cite{cracknell2014geological} integrated spatial coordinates as features for DNNs, while \cite{wang2019nearest} proposed a nearest neighbor neural network approach for geostatistical modeling. Others, such as \cite{zammit2022deep}, employed neural networks to estimate the warping function, enabling the transformation of the spatial domain to fit stationary and isotropic covariance structures. Convolutional neural networks have proven effective in capturing spatial dependencies in regularly gridded data \citep{shi2015convolutional, zammit2020deep, siden2020deep}. For irregularly gridded spatial data, diverse strategies have been proposed, including the use of basis functions as DNN features \citep{chen2020deepkriging}, decomposition of spatiotemporal processes into temporally referenced basis functions \citep{amato2020novel}, exploiting Gaussian Markov random fields with graph neural networks \citep{oskarsson2022scalable}, and the utilization of hierarchical statistical Integro-Difference Equation (IDE) frameworks \citep{zammit2020deep}. A comprehensive review is in \cite{wikle2023statistical}. 


The organizers of the KAUST Competition on Large Spatial Datasets 2023 simulated data at $N=10^5, 10^6$ spatial locations from zero-mean GPs with isotropic Mat\'ern correlation using the software \texttt{ExaGeoStat} \citep{abdulah2018exageostat}, divided the data into training and testing sets comprising of 90\% and 10\% of the spatial locations respectively and asked the participants to estimate the model parameters and predict the values at the testing sites along with uncertainty estimates. We compared several statistical and deep learning approaches using cross-validation, and we finally selected the Vecchia approximation technique \citep{vecchia1988estimation} for model fitting. This technique is implemented in the latest version of the \texttt{R} package \texttt{GpGp} \citep{guinness2021gpgp}. \textcolor{black}{From the perspective of the KAUST competition, it is worth noting that this package currently does not support the constraint of underlying GPs having a mean zero, and it does not return uncertainty estimates directly, except for the option of conditional sampling.} We developed additional \texttt{R} functions to bridge these gaps, along with implementing some further subsampling-based approximations and smoothing the skewed sampling distribution of the estimators in a parametric approach. \textcolor{black}{More details about the implementation are in Section \ref{subsec:GPGP}.} Additionally, for a large real spatial satellite-derived dataset on total precipitable water, we compared the predictive performances of some existing techniques based on multiple diagnostics.



The paper is organized as follows. In Section \ref{sec:competition}, we describe some necessary details about the 2023 KAUST competition on large spatial datasets. Section \ref{sec:statistical_models} briefly summarizes some statistical and deep learning models for analyzing large spatial datasets. Section \ref{sec:comparison_results} describes the results obtained based on fitting various statistical and deep learning approaches to the KAUST competition datasets. In Section \ref{sec:Data-Description}, we describe the results by comparing their prediction performances for a real dataset on total precipitable water. Section \ref{sec:conclusions} concludes.




\section{KAUST competition 2023 datasets}
\label{sec:competition}

The details about the competition are available \href{https://cemse.kaust.edu.sa/stsds/news/2023-kaust-competition-spatial-statistics-large-datasets}{here}. In this section, we briefly describe the structures of the training and testing datasets. For each of two choices of sample sizes (training plus testing) $N=10^5, 10^6$ in \eqref{eq:multivariate_normal}, the organizers simulated five datasets from zero-mean Gaussian processes with covariance structure \eqref{eq:matern_correlation} and divided each dataset into training and testing sets. The competition comprises four sub-competitions: (1a) Five training datasets each with $N_{\textrm{Train}}=90,000$ spatial locations, and the teams were asked to report the estimates of $\sigma^2$, $\phi$, $\nu$, and $\tau^2$ in \eqref{eq:matern_correlation}, along with 95\% confidence intervals, (1b) Similar to 1a except that $N_{\textrm{Train}}=900,000$, (2a) Corresponding to four training datasets in 1a, we have two testing datasets each of dimension $N_{\textrm{Test}}=10,000$ and for one dataset, only one testing dataset of the same $N_{\textrm{Test}}$ is available, where the teams were asked to submit predictions, along with 95\% prediction intervals, and (2b) Similar to 2a except that $N_{\textrm{Train}}=900,000$ and $N_{\textrm{Test}}=100,000$.

\begin{figure}
    \centering
    \includegraphics[height = 0.25\linewidth]{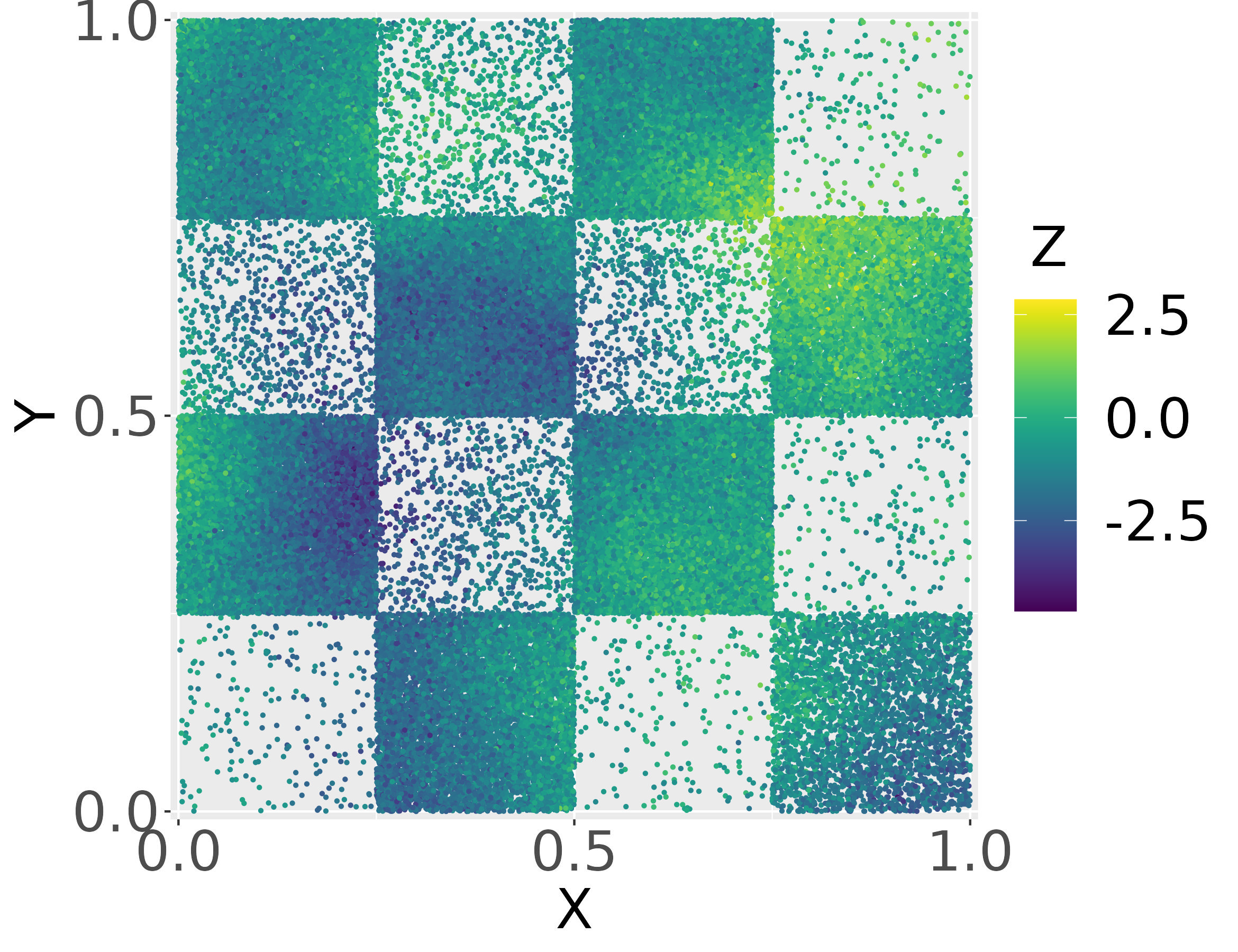}
    \includegraphics[height = 0.25\linewidth]{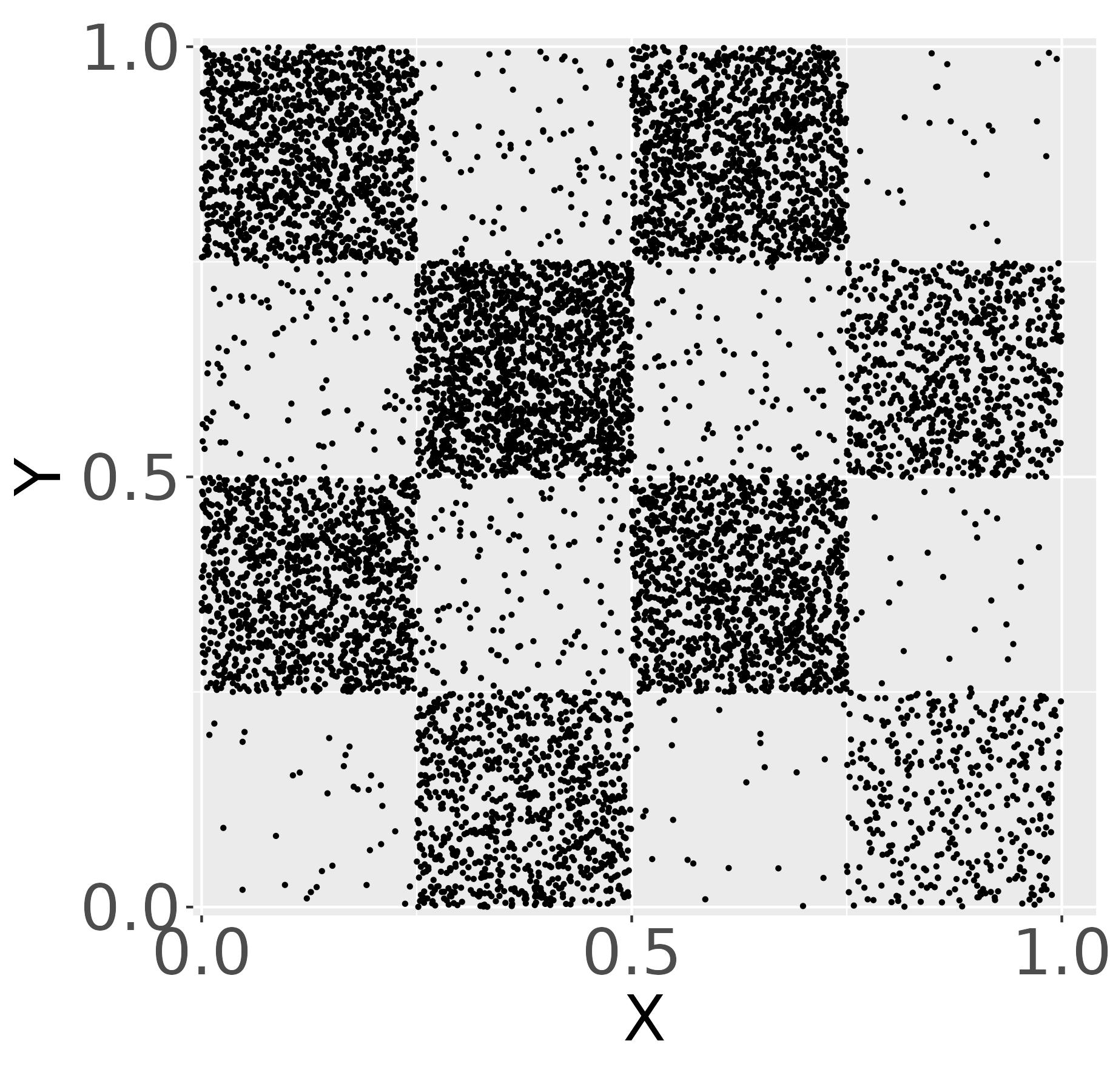}
    \includegraphics[height = 0.25\linewidth]{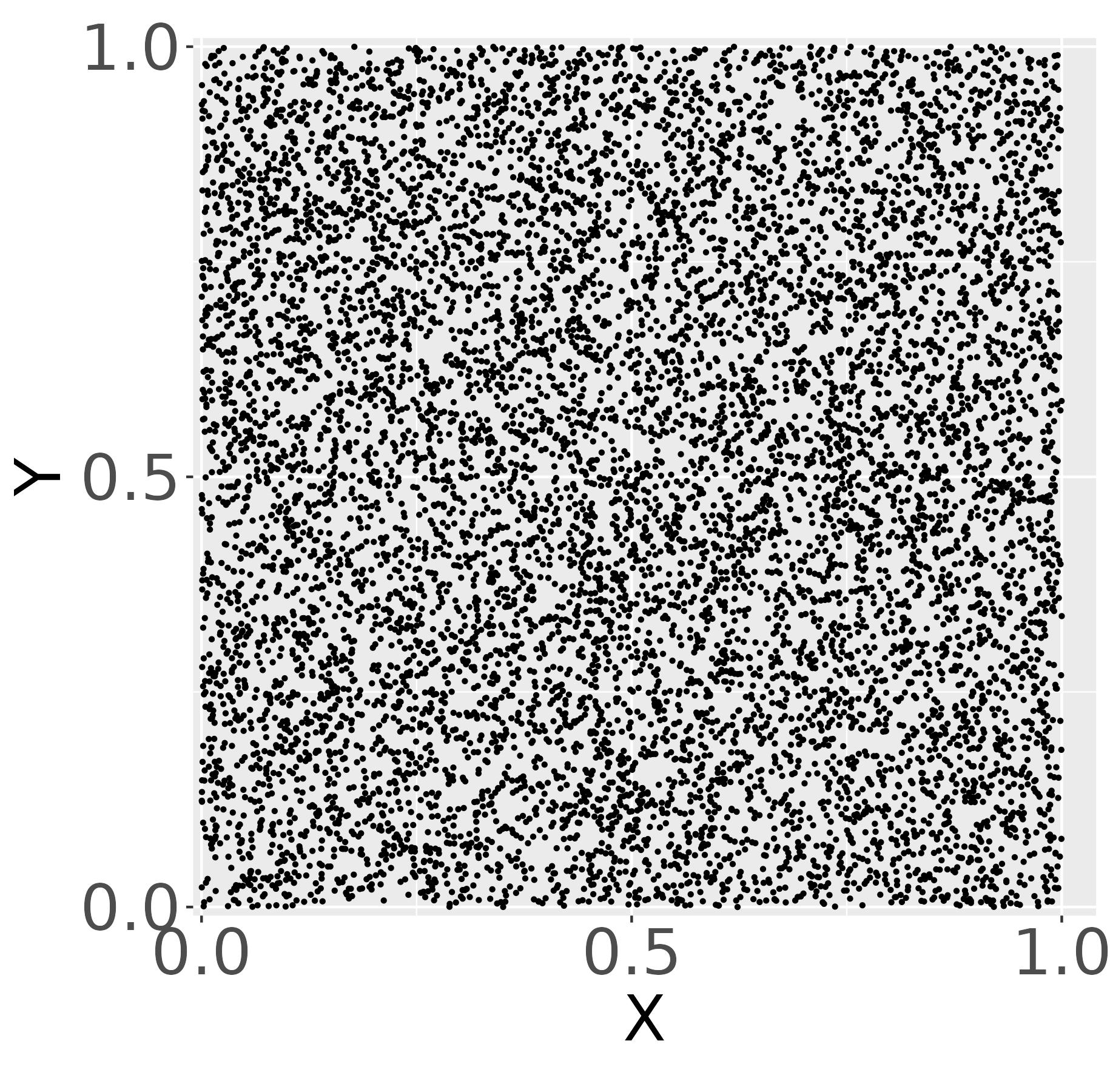}

\includegraphics[height = 0.25\linewidth]{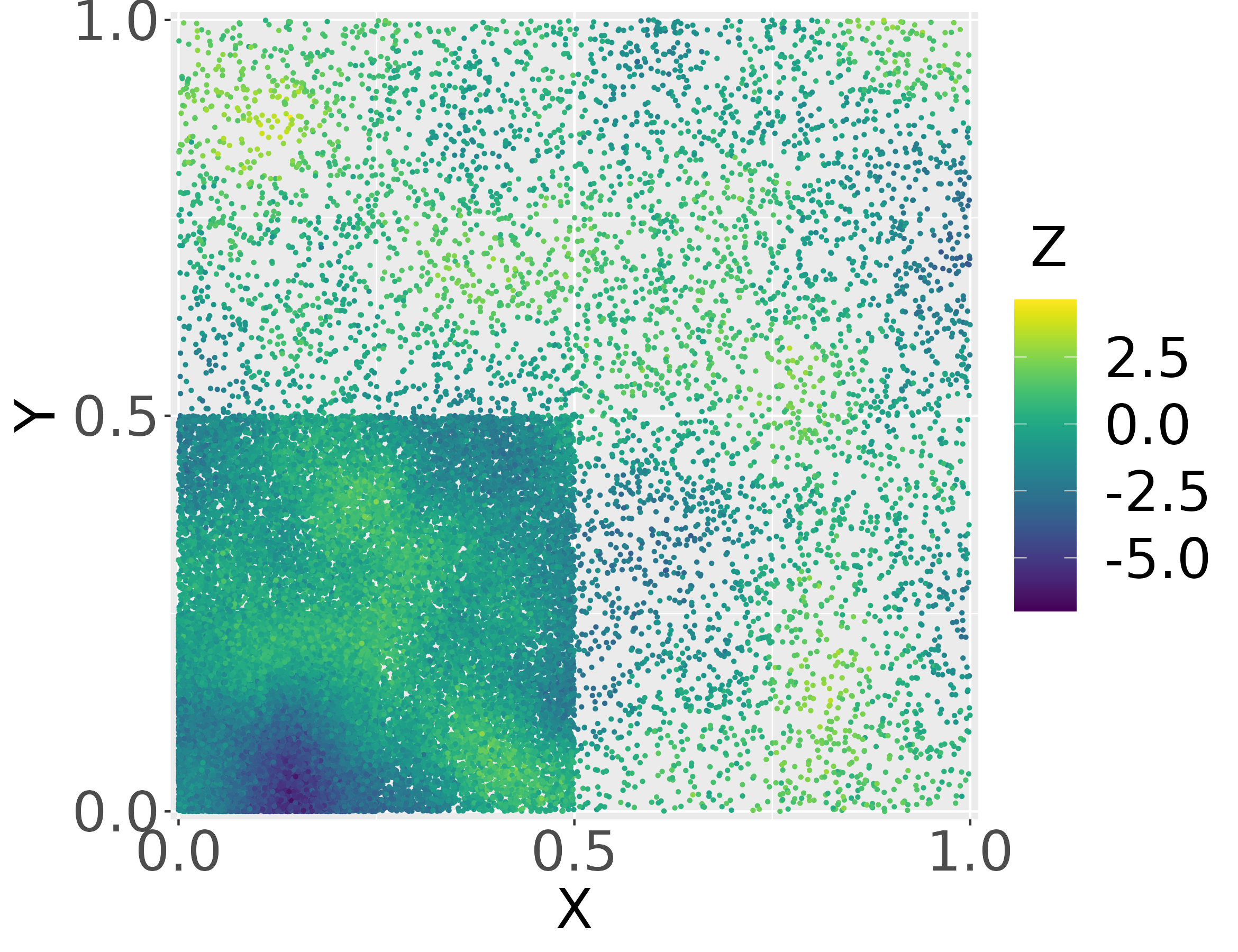}
    \includegraphics[height = 0.25\linewidth]{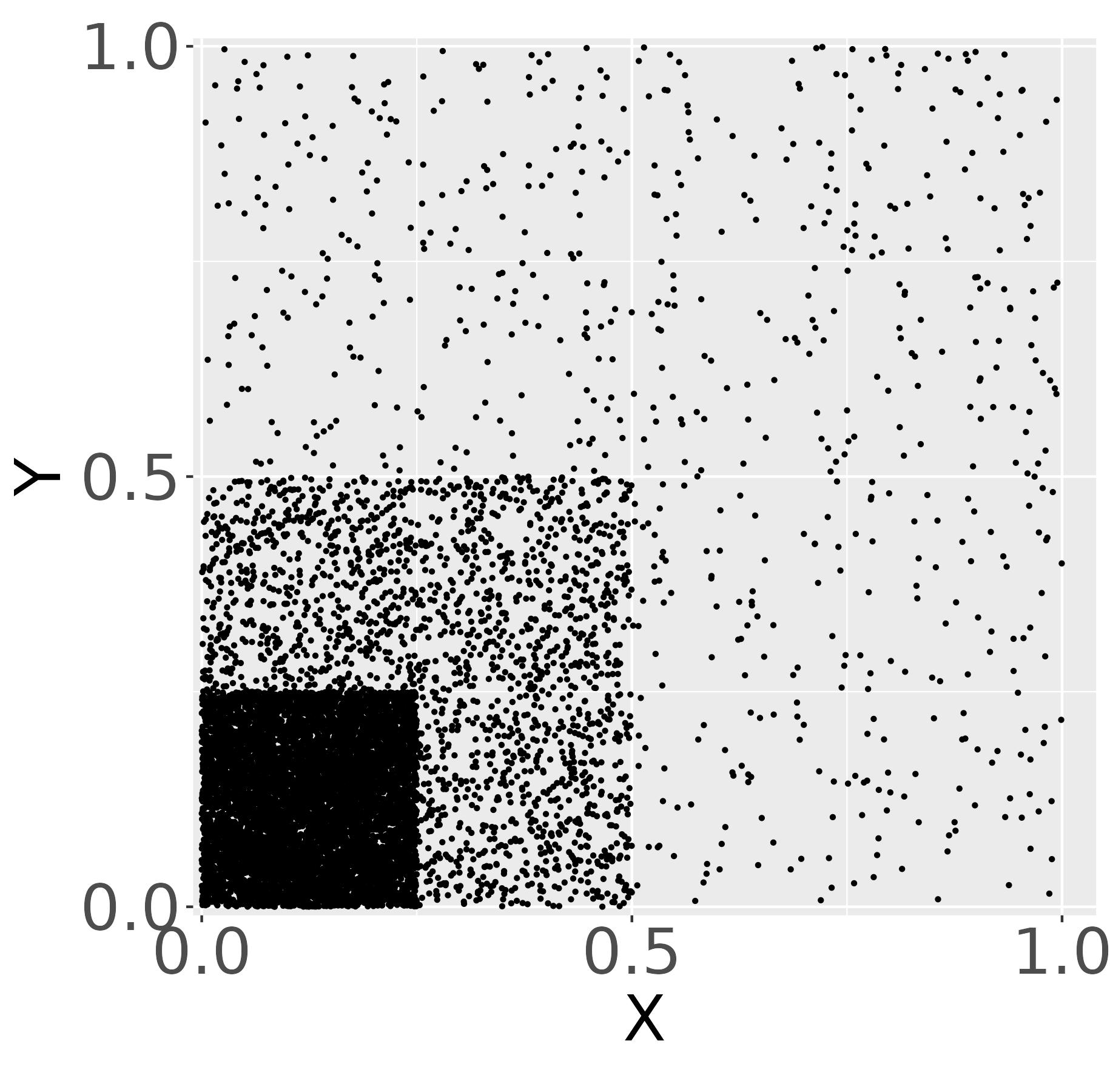}
    \includegraphics[height = 0.25\linewidth]{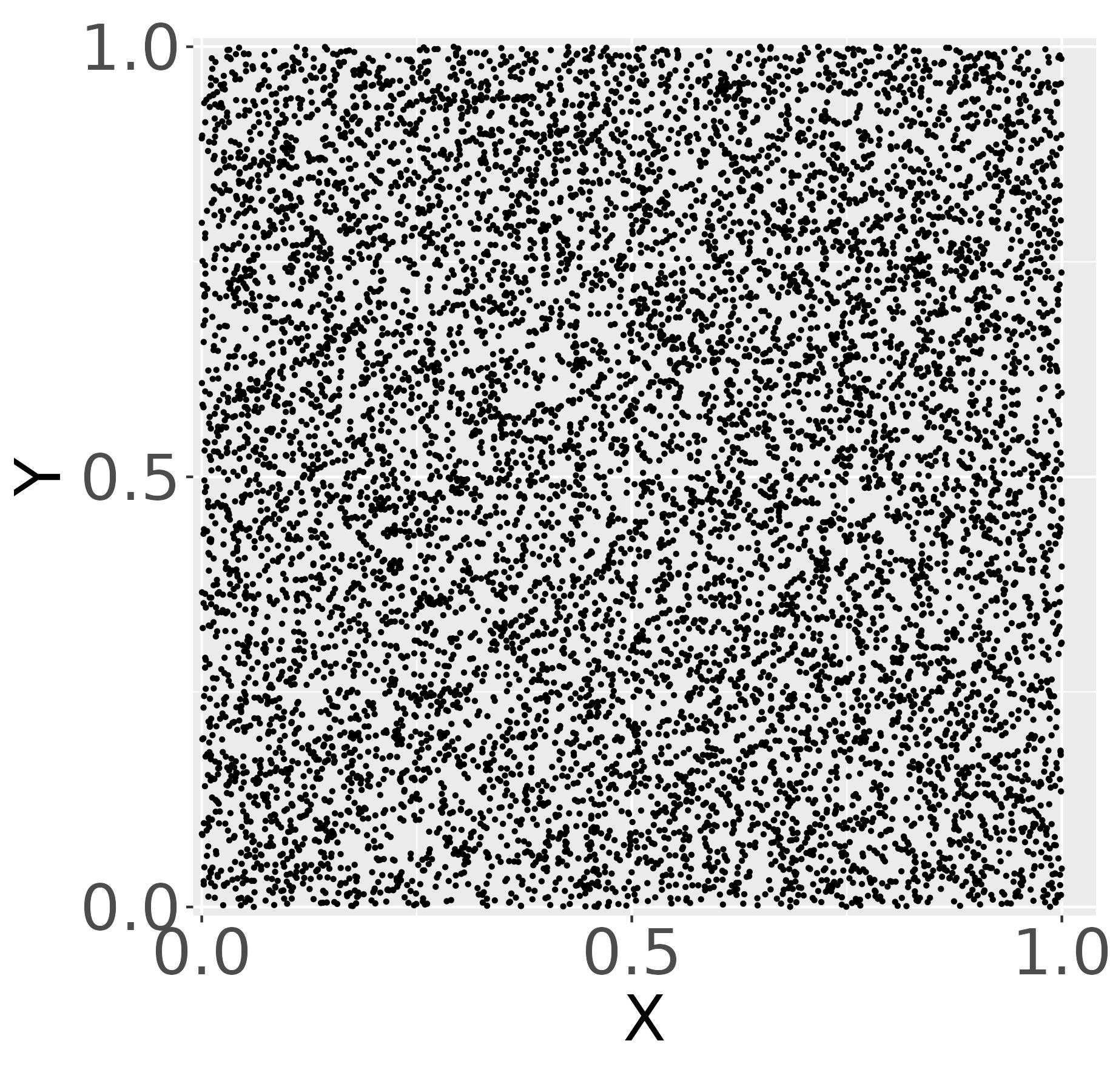}

\includegraphics[height = 0.25\linewidth]{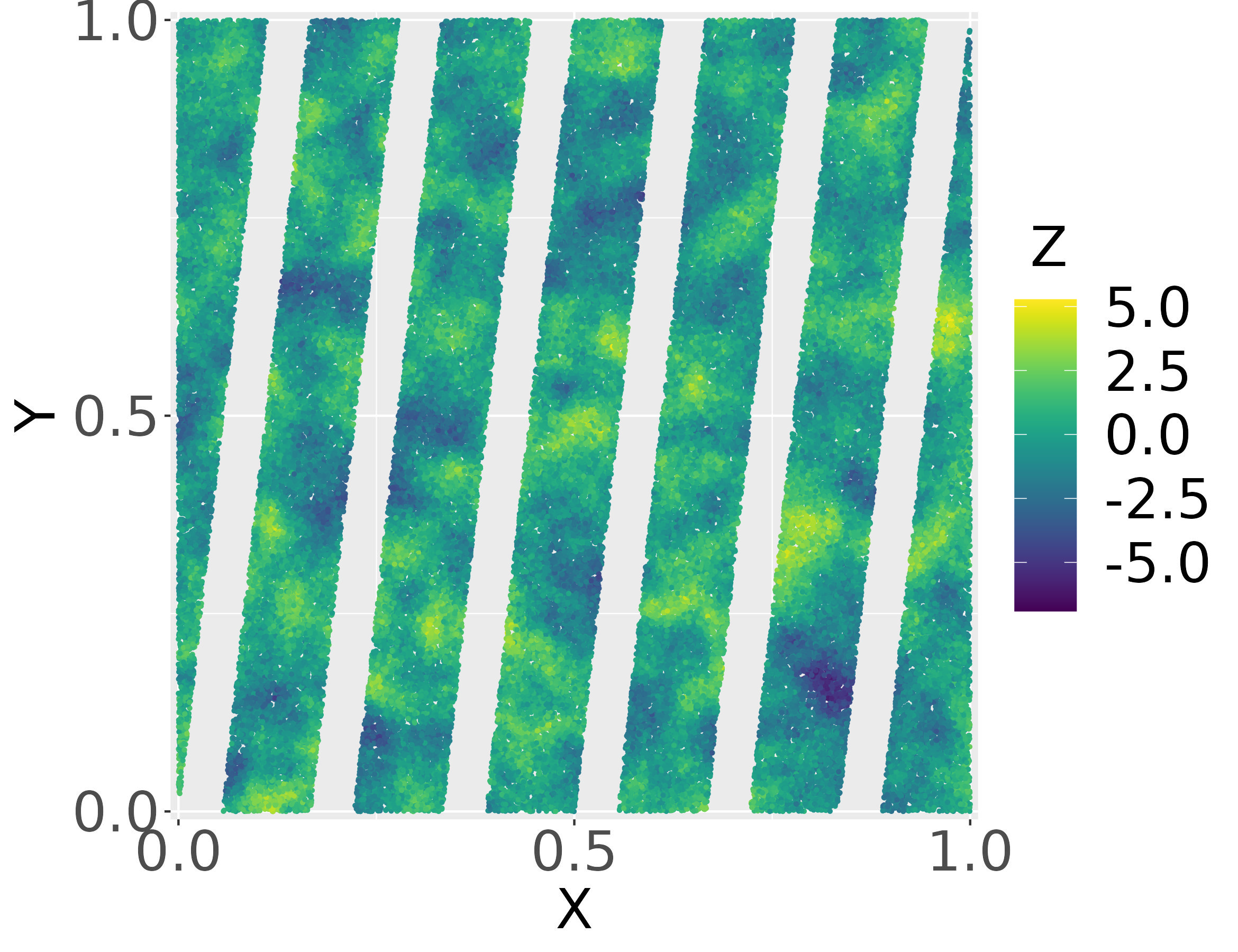}
    \includegraphics[height = 0.25\linewidth]{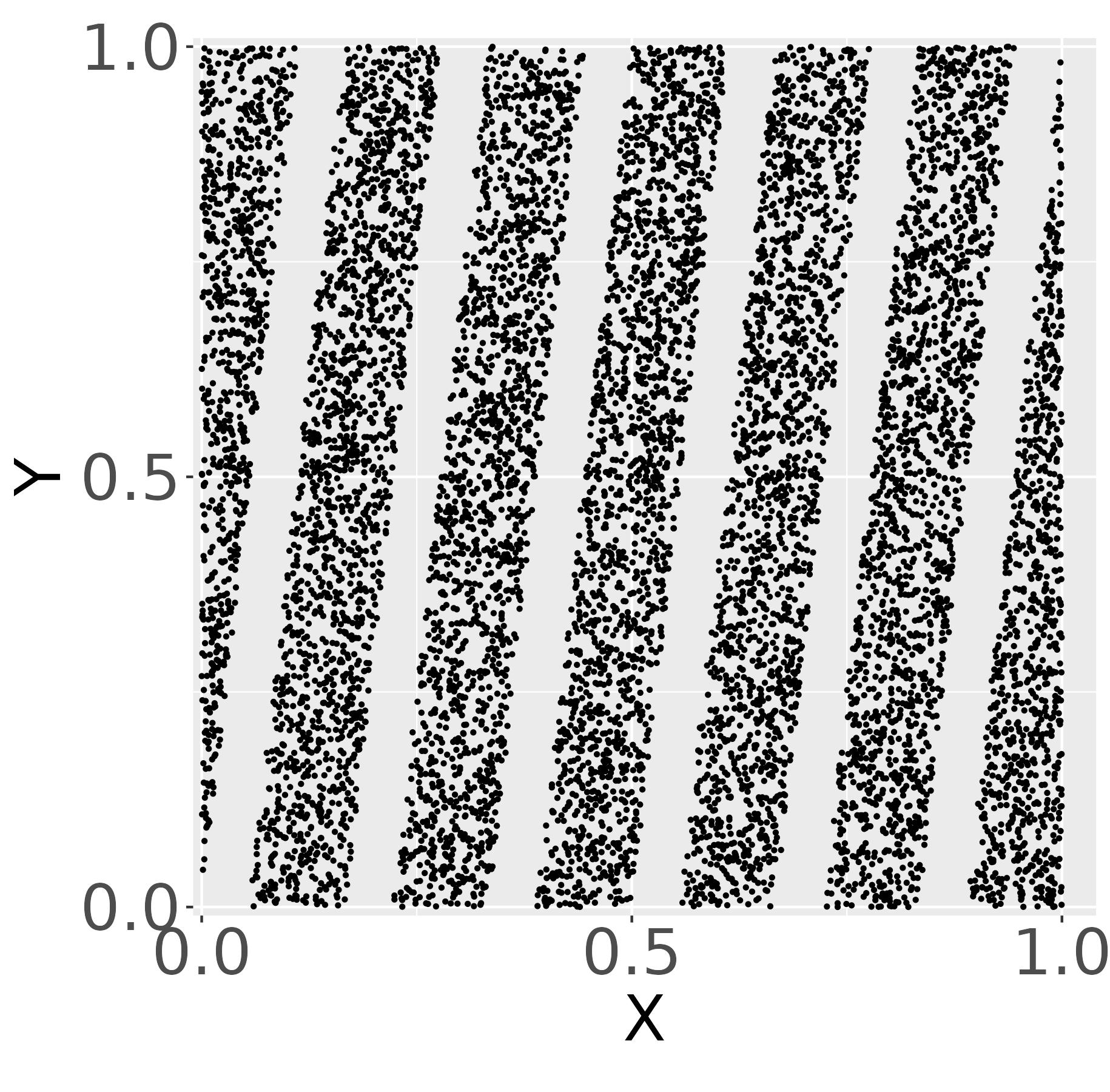}
    \includegraphics[height = 0.25\linewidth]{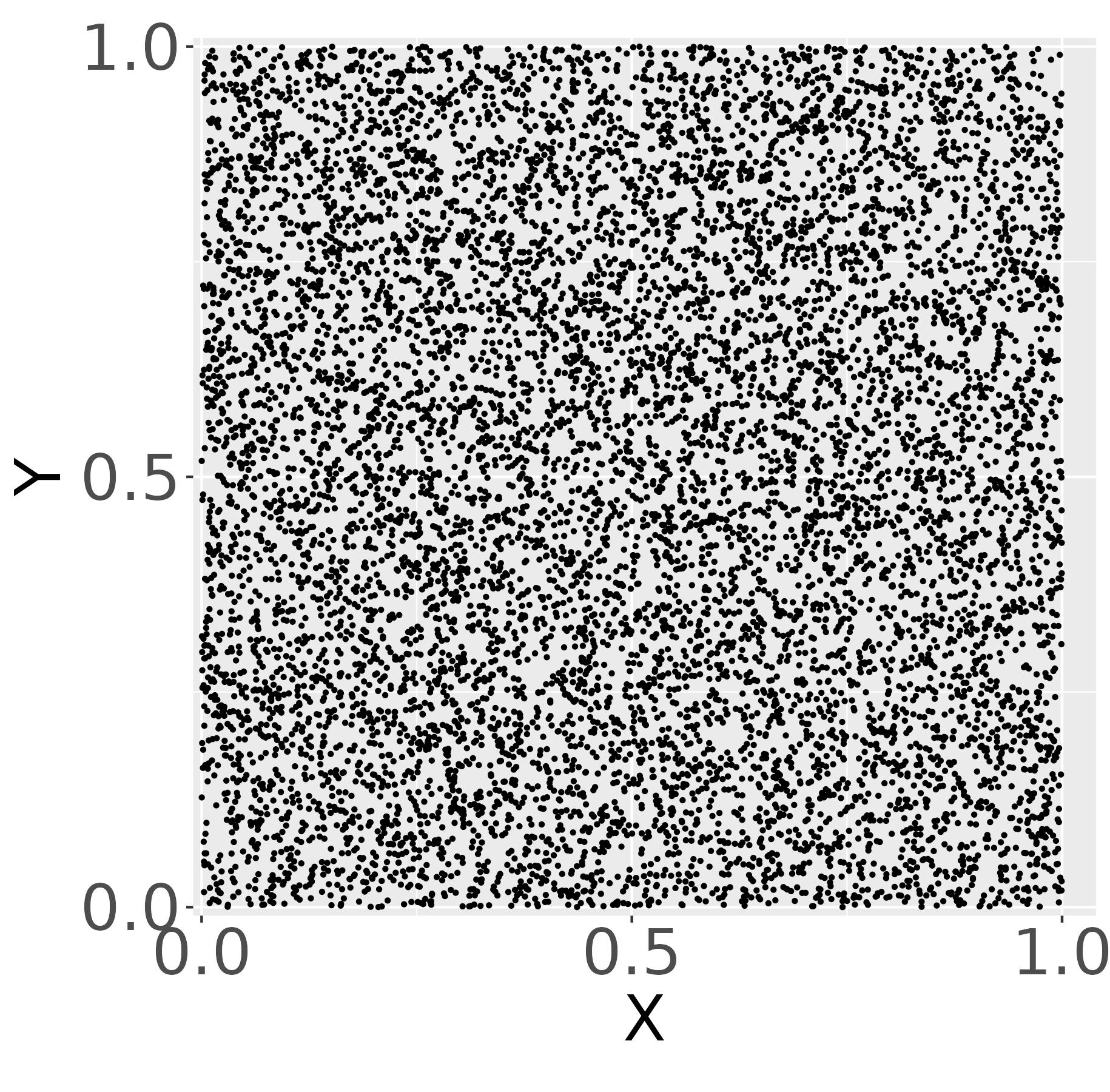}

\includegraphics[height = 0.25\linewidth]{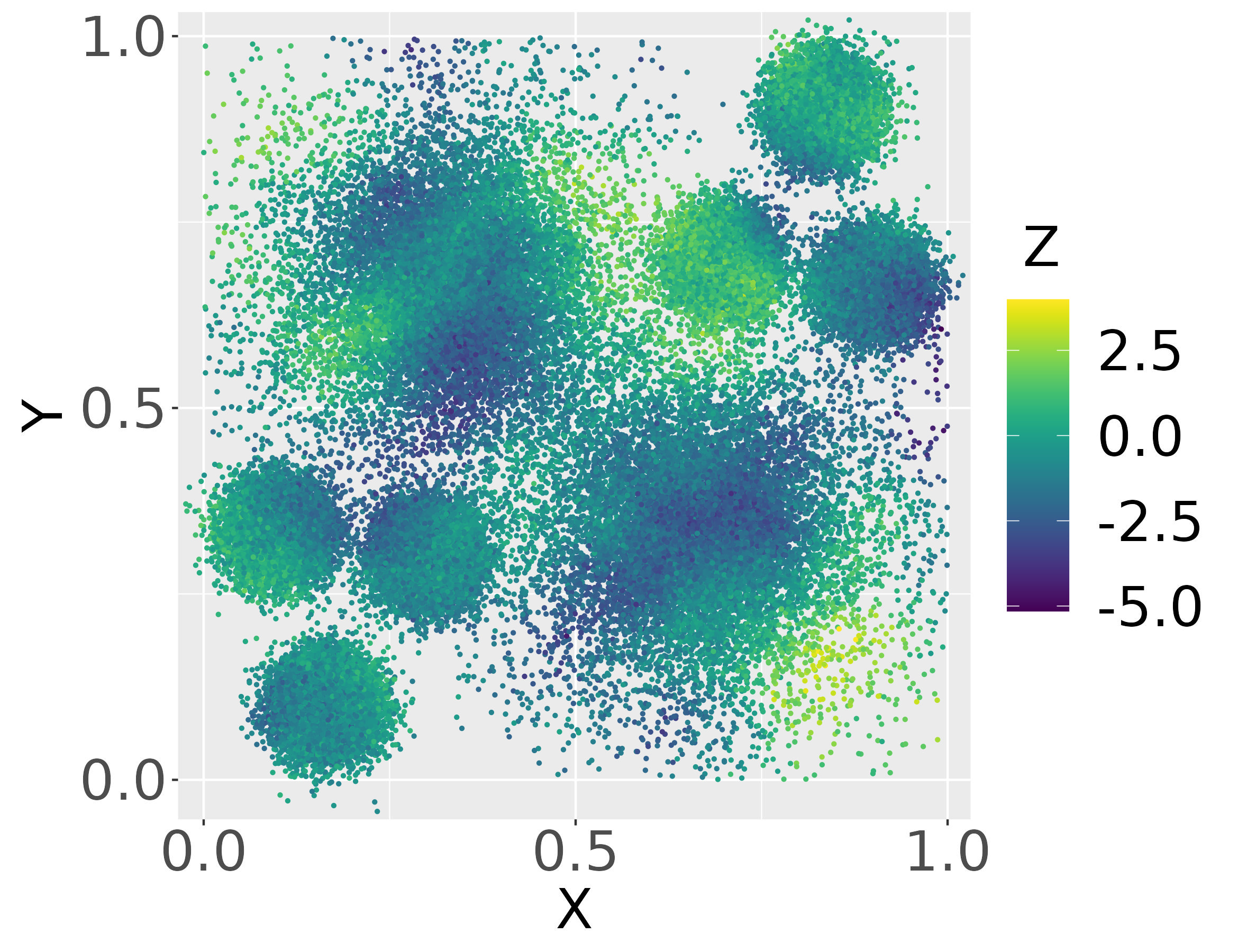}
    \includegraphics[height = 0.25\linewidth]{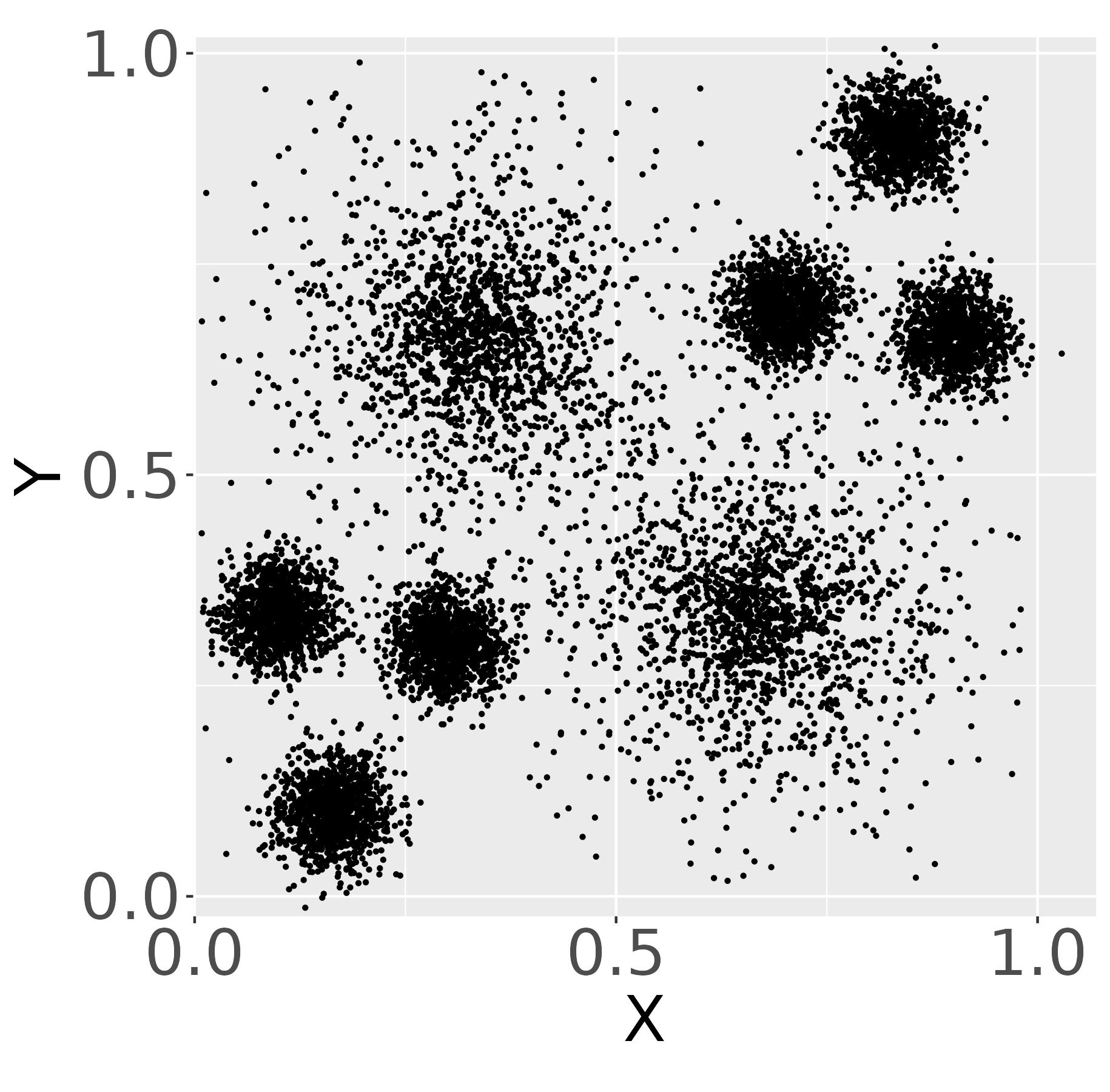}
    \includegraphics[height = 0.25\linewidth]{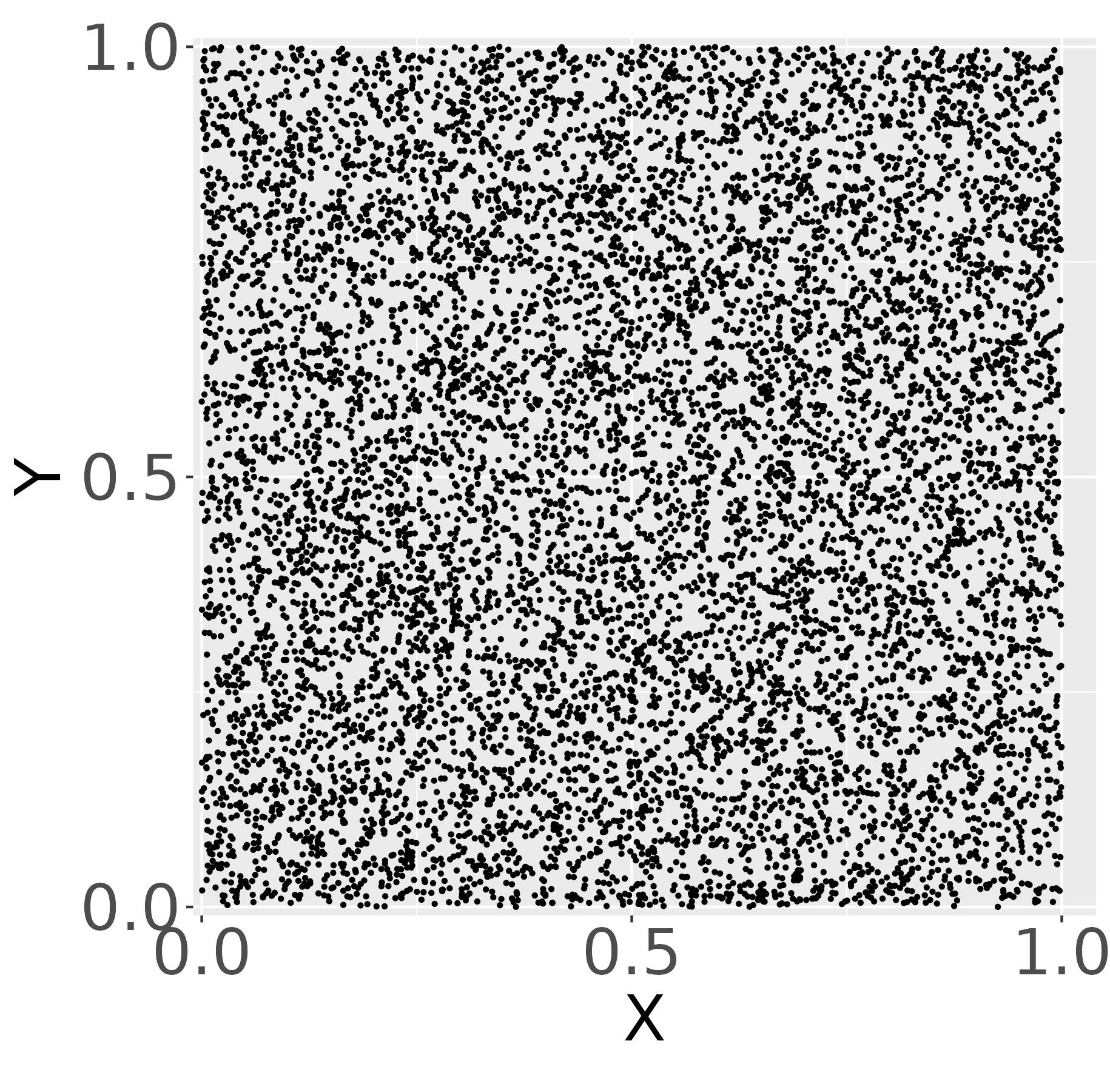}

\includegraphics[height = 0.25\linewidth]{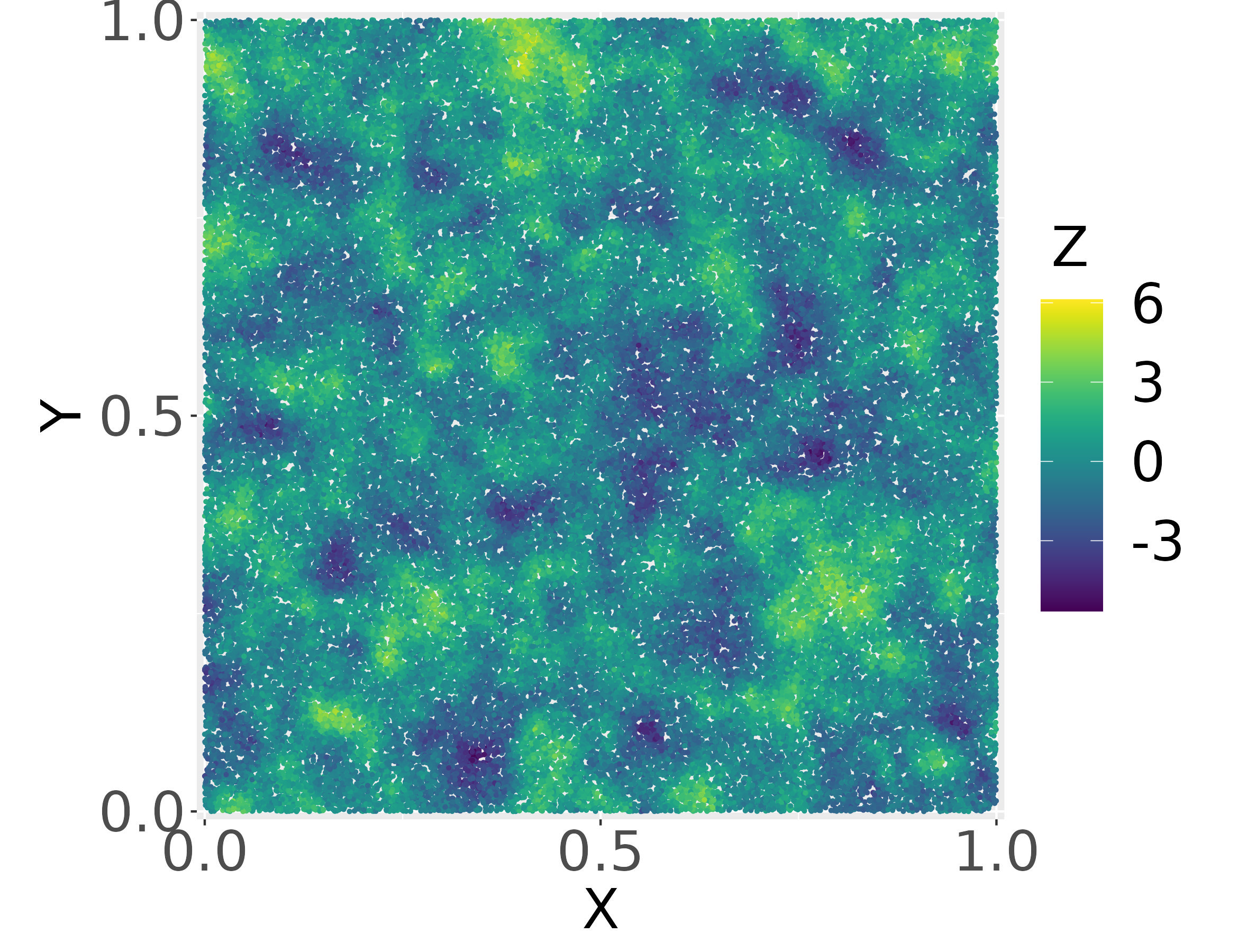}
\hspace{4.5cm}
    \includegraphics[height = 0.25\linewidth]{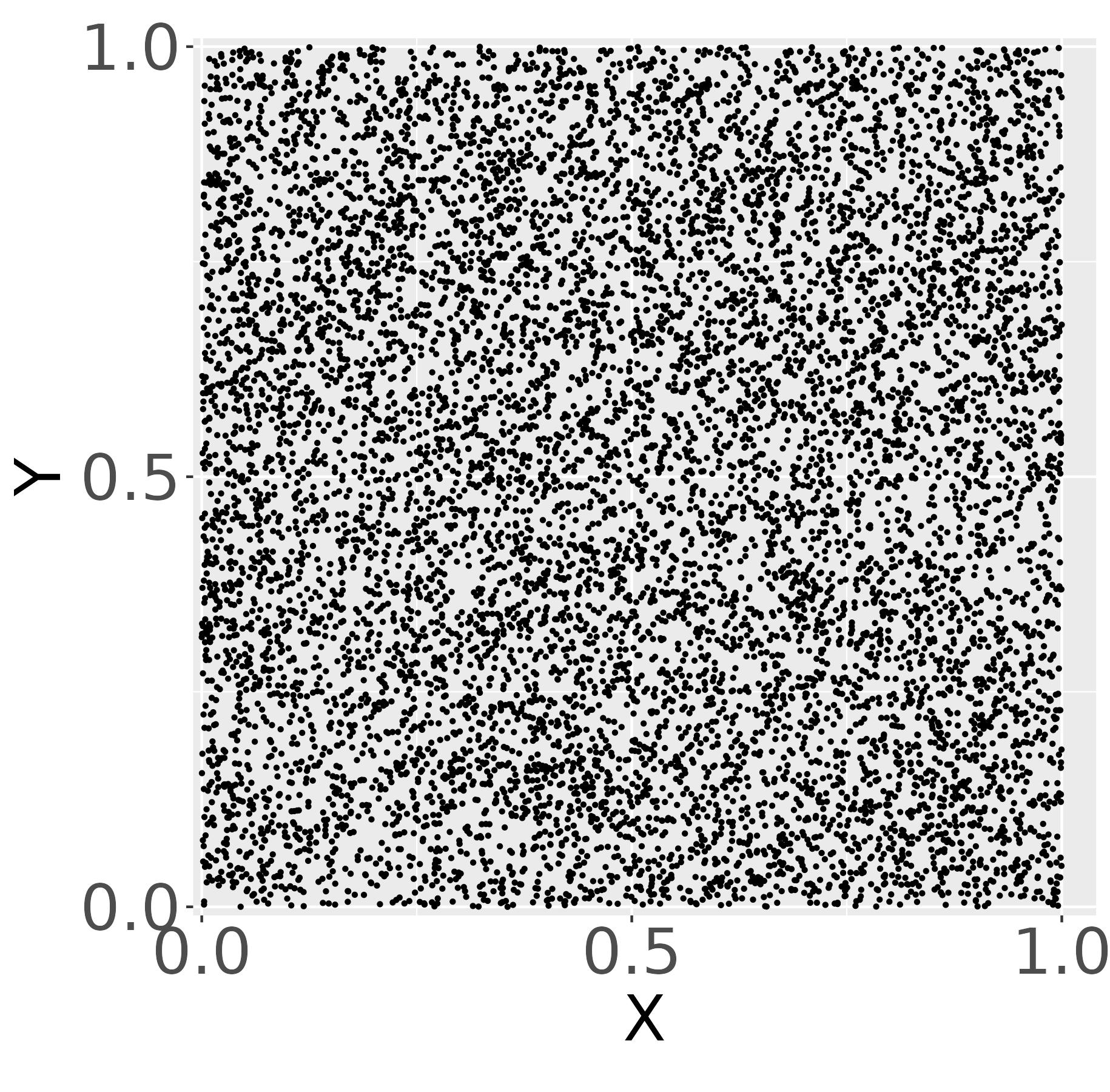}
    \caption{Five training datasets (left column) and the corresponding testing datasets (second and third columns) for Sub-competitions 1a and 2a. For the fifth training dataset (bottom row), only one corresponding testing dataset is provided. The colors in the first column represent the realizations from different Gaussian processes.}
    \label{fig:train_test_maps}
\end{figure}

The training and corresponding testing datasets for Sub-competition 1a/2a are presented in Figure \ref{fig:train_test_maps}. For the first training dataset, most locations are distributed densely within 50\% area of the domain; this indicates a non-homogeneous point pattern. The realizations show a smooth sample path indicating a high spatial range. For the second training dataset, only 6.25\% area has a very dense spatial distribution of observation locations, a medium dense spatial point pattern is observable within 18.75\% area, and the rest of the 75\% region has a very sparse distribution of spatial points. Similar to the first training dataset, the smooth sample path again indicates a high spatial range. The third training dataset does not contain data locations over six spatial stripes resembling the situation of satellite paths. The spatial sample path appears to be more jittery than the two previous datasets, and thus, a smaller value of the range parameter is expected. For the fourth dataset, most training data locations concentrate near six small circular regions, and the rest spread across two large circular regions. As a result, no training data are available at a large portion of the spatial domain. A smoother sample path is observable compared to the third dataset indicating a high value of the range parameter. The fifth training dataset shows a homogeneous spatial point pattern. The jittery sample path indicates a small spatial range.

The point patterns of the first set of testing datasets resemble the patterns of the corresponding training datasets, whereas the second set of testing datasets shows a homogeneous spatial point pattern. Here, prediction uncertainty for the first set is expected to be lower than that for the second set due to the data availability in the neighborhood; for the second set, there is either no information available within a small neighborhood (third and fourth training datasets) or relatively less information available in a significantly large portion of the spatial domain (first and second training datasets). 

\begin{figure}[h]
    \centering
    \includegraphics[height = 0.19\linewidth]{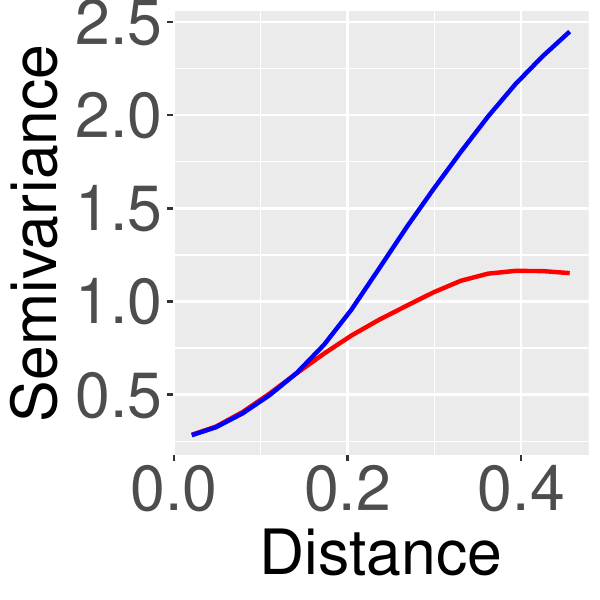}
    \includegraphics[height = 0.19\linewidth]{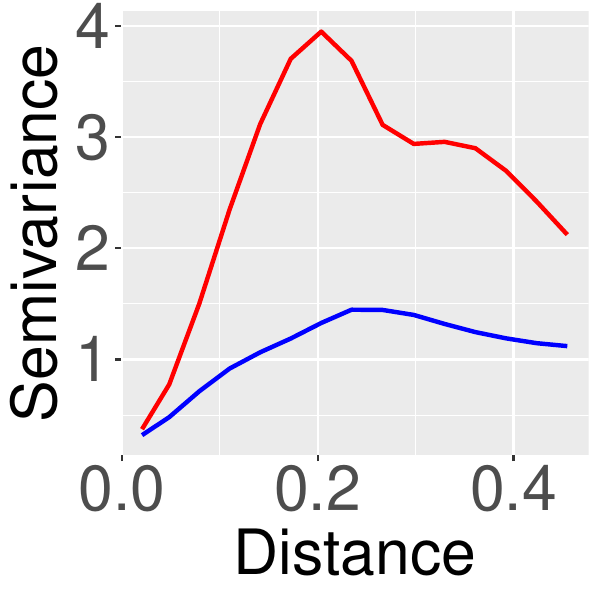}
    \includegraphics[height = 0.19\linewidth]{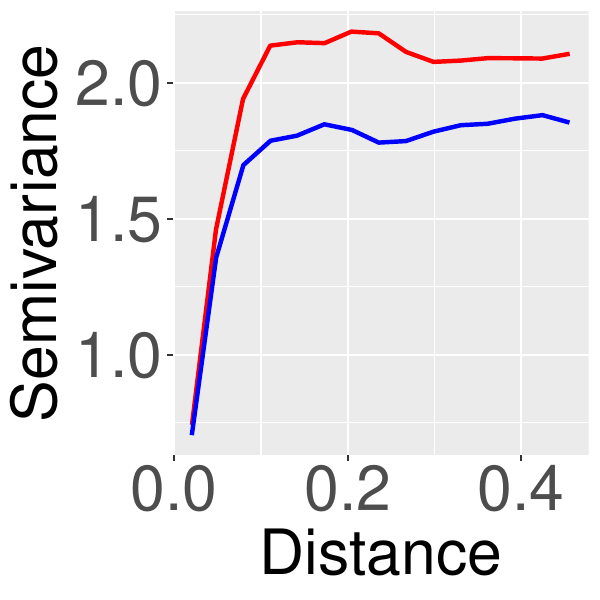}
    \includegraphics[height = 0.19\linewidth]{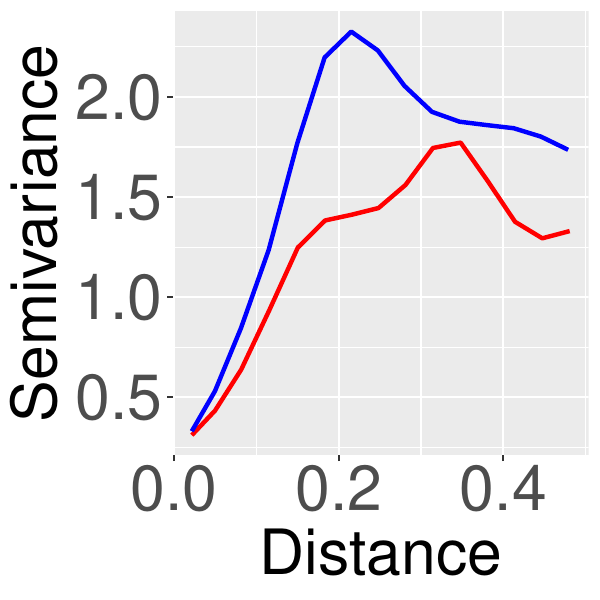}
    \includegraphics[height = 0.19\linewidth]{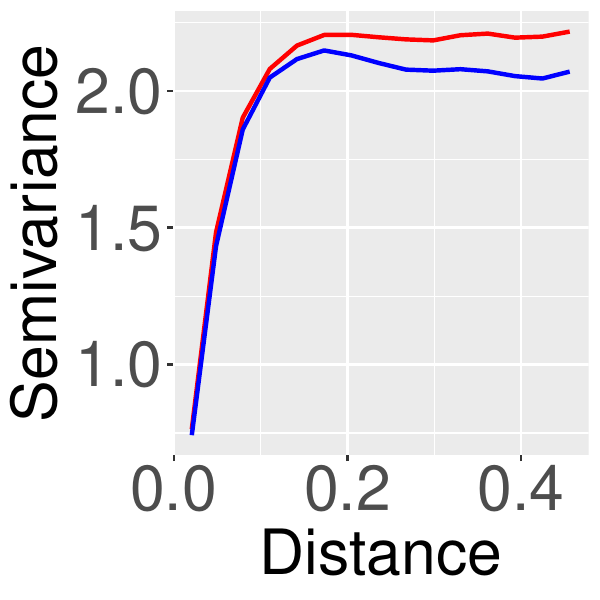}
    \caption{Empirical semivariance profiles for the five training datasets, with red and blue lines representing the corresponding datasets from Sub-competition 1a and 1b, respectively.}
    \label{fig:semivariance_emp}
\end{figure}

For Sub-competitions 1b and 2b, the spatial point patterns of the training and testing datasets are similar to that of 1a and 2a and hence not presented. For training datasets of Sub-competitions 1a and 1b, we show the empirical variograms in Figure \ref{fig:semivariance_emp}. The nugget component appears small and similar for all datasets, while the partial sill varies across datasets and sub-competitions. Compared to other datasets, the spatial range appears smaller for the third and fifth ones; our observations from Figures \ref{fig:train_test_maps} and \ref{fig:semivariance_emp} thus coincide.


\section{Statistical and deep learning models for large spatial datasets}
\label{sec:statistical_models}
Let $Y(\bm s)$ be the response variable of interest at a location $\bm s$ as in \eqref{eq:multivariate_normal}, and $\mathbf{X}(\bm s)$ is a $P$-length vector of covariates observed at $\bm s$.  If the covariates fail to capture all the structured spatial variation of the response, the spatial linear mixed effects (SLME) model \citep{banerjee2003hierarchical,cressie2015statistics}  for analyzing the response over $\mathcal{D} \subset \mathbb{R}^2$ can be written as
\begin{align}
\label{eq:mixedeffects.model}
Y(\bm s)=\mathbf{X}(\bm s)^{'} \bm\beta+W(\bm s)+\varepsilon(\bm s),~~ \bm{s} \in \mathcal{D},
\end{align}
where $\bm \beta$ is the $P$-length vector of regression coefficients, $\varepsilon(\bm s)$'s are Gaussian white noises that capture the small scale and unstructured variation in the data and they have mean $0$ and variance $\tau^2$, and $W(\bm s)$'s are spatial random effects that are assumed to follow a zero-mean Gaussian process (GP) with Mat\'ern covariance function \eqref{eq:matern_correlation} with $\tau^2 = 0$. As mentioned in Section \ref{sec:introduction}, the trend component $\mathbf{X}(\bm s)^{'} \bm\beta $ may be easily handled using regression techniques leaving the main computational burden to the dependence-related component $\tilde{W}(\bm s) = W(\bm s) + \varepsilon(\bm s)$; in case realizations from $Y(\cdot)$ are available at $N$ spatial locations, the corresponding covariance matrix $\bm{\Sigma}_{\bm{\theta}}$, constructed following \eqref{eq:matern_correlation} with $\bm{\theta} = (\sigma^2, \phi, \nu, \tau^2)'$, would be a $(N \times N)$-dimensional dense matrix, and for likelihood-based inference, we need to evaluate \eqref{eq:multivariate_normal} which involves computing costly determinant and inverse of $\bm{\Sigma}_{\bm{\theta}}$, when $N$ is large. Many approximation methods have been developed to improve the computational efficiency; it is now possible to fit the model \eqref{eq:mixedeffects.model} for large spatial datasets \citep{sun2012geostatistics}. Alternatively, deep learning-based models have recently gained attention in the spatial prediction problem that often utilizes the spatial basis functions as an input to model any nonlinear function of type \eqref{eq:mixedeffects.model}. Despite their black-box nature, their flexibility and handling of vast amounts of data are well-appreciated. Below we provide brief details about some of the well-known frequentist (Section \ref{subsec:Frequentist.methods}), Bayesian (Section \ref{subsec:Bayesian.methods}),  and deep learning models (Section \ref{subsec:deepL.methods}) that provide computational bottleneck for large datasets; see also \cite{heaton2019case} for a brief overview of statistical models and \cite{zammit2022deep} for deep learning models for large-scale estimation and spatial prediction problems.


\subsection{Frequentist approaches}
\label{subsec:Frequentist.methods}
The SLME model \eqref{eq:mixedeffects.model} may be analyzed using the frequentist criterion where the full log-likelihood is given as
\begin{align}
\label{eq:Freq.logilk}
\ell(\bm{\theta}, \bm \beta; \bm{y}) =\log f_{\bm \theta, \bm \beta}(\bm y) = -\ffrac{1}{2}\log \left( \vert 2\pi \bm{\Sigma}_{\bm{\theta}} \vert \right) -\ffrac{1}{2} \left(\bm{y}- \mathbf{X}\bm \beta \right)' \bm{\Sigma}_{\bm{\theta}}^{-1} \left(\bm{y}- \mathbf{X}\bm \beta \right),
\end{align}
where $\bm y=[y(\bm s_1),\ldots,y(\bm s_N)]'$ are the realizations of the process $Y(\cdot)$, $\mathbf{X}$ is the $N\times P$ design matrix with its $i$-th row being $\mathbf{X}(\bm{s}_i)$, $\bm{\Sigma}_{\bm{\theta}}$ is the covariance matrix, and $f_{\bm \theta, \bm \beta}(\bm y):= \mathcal{N}_N(\mathbf{X} \bm \beta,\bm \Sigma_{\theta})$ denote the $N$-variate Gaussian density in \eqref{eq:multivariate_normal} with mean $\bm{\mu} = \mathbf{X} \bm \beta$ and covariance matrix $\bm{\Sigma} = \bm{\Sigma}_{\bm{\theta}}$. For $N \approx 10^5$, optimizing this likelihood function is infeasible due to the involved computational burden, and some approximate methods must be adopted to optimize \eqref{eq:Freq.logilk}. Below we detail a few frequentist approaches that circumvent this issue.



\subsubsection{Vecchia approximation (\texttt{GpGp}, \texttt{GpGp0})} 
Vecchia’s approach \citep{vecchia1988estimation} is one of the earliest proposed GP approximations, which leads to a sparse Cholesky factor of the precision matrix $\bm{\Sigma}_{\bm{\theta}}^{-1}$. Based on some ordering of the observation locations, Vecchia’s approximation replaces \eqref{eq:multivariate_normal} with a product of univariate conditional Gaussian densities, where each conditional distribution conditions on only a small subset of previous observations in the ordering. More explicitly, let $g(1)=\emptyset$, $g(i) \subseteq \{1,\ldots,i-1\}$, and $\bm y_{g(i)}$ be the corresponding subvector of $\bm y$. Then, Vecchia’s approximation of \eqref{eq:Freq.logilk} is given by
\begin{align}
\label{eq:vecchia-logLik}
    \ell_v(\bm{\theta}, \bm \beta; \bm{y}) =  \sum_{i=1}^{N} \log f_{\bm{\theta}, \bm \beta}\left(y_i \mid \bm y_{g(i)}\right),
\end{align}
where $f_{\bm{\theta}, \bm \beta}\left(y_i \mid \bm y_{g(i)}\right)$ denote the generic univariate conditional Gaussian density. This approximation incurs a low computational and memory burden whenever the cardinality of $g(i)$ is small and corresponds to a valid multivariate normal distribution with the same mean as the original process and a covariance matrix that approximates the true $\bm{\Sigma}_{\bm{\theta}}$ very accurately in terms of Kullback–Leibler divergence \citep{katzfuss2021general}. Reordering and grouping of observations is one of the key challenges in Vecchia’s approximation, and to deal with it, \cite{guinness2018permutation} proposed an automatic grouping and reordering method that improves the approximation accuracy drastically. Furthermore, \cite{guinness2021gaussian} used the Fisher scoring rule for a fast and reliable way of maximizing the log-likelihood \eqref{eq:vecchia-logLik}. 



The original idea of Vecchia’s approximation \citep{vecchia1988estimation}, together with reordering and grouping \citep{guinness2018permutation} and Fisher scoring rule \citep{guinness2021gaussian}, is implemented in \texttt{R} package \texttt{GpGp} \citep{guinness2021gpgp}. The main function for parameter estimation in this package \texttt{fit\textunderscore model} assumes the default data structure as in \eqref{eq:mixedeffects.model} without allowing the user to set the mean to zero. For the KAUST competition, where the participants were informed that the datasets were simulated from zero-mean GPs, we developed an additional function \texttt{fit\textunderscore model\textunderscore meanzero}; this function allows the user to set the mean process to zero. We compared the performances using both functions; see Section \ref{subsec:GPGP} for further details. Henceforth, we use the terms \texttt{GpGp} and \texttt{GpGp0} to denote the Vecchia’s approximation implemented by \cite{guinness2021gpgp} and our modified version allowing zero mean, respectively. In both cases, a single core was used for computation.



\subsubsection{Maximum Block Composite Likelihood estimation and  prediction (\texttt{BCL})} 
The block composite likelihood \citep{eidsvik2014estimation} is built from the joint densities of pairs of neighboring spatial blocks that allow a large dataset to split into smaller datasets which can then be evaluated separately before combining them through a product. More explicitly, we divide the spatial domain into blocks by a Voronoi tessellation of the data coordinates, with cells extending for all depths. The tessellation is made by random sampling, without replacement, among all data sites. We form a partition of $\tilde{N}$ blocks $\{\mathcal{B}_1, \ldots, \mathcal{B}_{\tilde{N}}\}$ such that $\cup_{i=1}^{\tilde{N}}\mathcal{B}_i = \mathcal{S}$ and $\mathcal{B}_i \cap \mathcal{B}_j = \O$ for $i \neq j$. In our application, we assume independence across blocks; however, certain spatial Markov structures can further be imposed across blocks. Thus, denoting the data vector corresponding to partition $\mathcal{B}_i$ by $\bm{y}_i$ with design matrix $\mathbf{X}_i$, and the corresponding covariance matrix by $\bm{\Sigma}^{(i)}_{\bm{\theta}}$, the log-likelihood \eqref{eq:Freq.logilk} is approximated by 
\begin{align}
\label{eq:BCL_logLIk}
\ell^*(\bm{\theta},\bm \beta; \bm{y}_1, \ldots, \bm{y}_{\tilde{N}}) = -\ffrac{1}{2} \sum_{i=1}^{\tilde{N}}\log(\vert 2\pi \bm{\Sigma}_{\bm{\theta}}^{(i)} \vert) -\ffrac{1}{2} \sum_{i=1}^{\tilde{N}}(\bm{y}_i - \mathbf{X}_i\bm \beta)' \bm{\Sigma}_{\bm{\theta}}^{(i)-1}(\bm{y}_i - \mathbf{X}_i\bm \beta),
\end{align}
which is then maximized to obtain the maximum likelihood estimates of $\bm{\theta}$ and $\bm\beta$.


Here the advantage is that the summation step over $\{\mathcal{B}_1, \ldots, \mathcal{B}_{\tilde{N}}\}$ can be implemented in parallel and we used 20 cores in our application. For our implementation, we set $\tilde{N} = \lfloor N/500 \rfloor$. For kriging, one simple approximation approach is to use ordinary kriging method within a small neighborhood; we chose a square region $(s_X^*-\Delta, s_X^* + \Delta) \times (s_Y^*-\Delta, s_Y^* + \Delta)$ for a prediction location $\bm{s}^* = (s_X^*, s_Y^*)'$, where the choice of $\Delta$ is problem specific. If more than 500 observation locations lie within the neighboring square, we kept only the nearest 500 observations in our implementation to speed up the computation. The kriging across the prediction locations can be performed in parallel and we distributed it over 20 cores. We set the mean to zero and only optimized $\bm{\theta}$ in \eqref{eq:BCL_logLIk}.

\subsubsection{\bf Fixed Rank Kriging (\texttt{FRK})}
Fixed Rank Kriging \citep{cressie2008fixed} is a spatial prediction strategy built on the concept of the SLME model \eqref{eq:mixedeffects.model}, also known as the spatial random effects (SRE) model. 
The primary assumption of FRK is that the spatial process $W(\cdot)$ can be approximated
by a linear combination of $K$ spatial basis functions $\{h_1(\cdot), \ldots, h_K(\cdot)\}$ and $K$ random basis-function coefficients $\{A_1, \ldots, A_K\}$. The use of $K$ basis functions ensures that all estimation and prediction equations only
contain inverses of matrices of size $K \times K$, where $K << N$. In practice, the set $\{h_1(\cdot), \ldots, h_K(\cdot)\}$ is comprised of functions at $R$ different resolutions such that the approximated process can be written as
$$\overline{W}(\bm{s}) = \sum_{r=1}^R \sum_{k=1}^{K_r} h_{rk}(\bm{s}) A_{rk} \approx W(\bm{s}), ~~ \bm{s} \in \mathcal{D},$$
where $h_{rk}(\bm{s})$ is the $k$-th spatial basis function at the $r$-th resolution with associated coefficient $A_{rk}$, and $K_r$ is the number of basis functions at the $r$-th resolution, such that $K = \sum_{r=1}^R K_r$. We used $R = 2$ resolutions of compactly supported bisquare basis functions and a total of $K \approx 150$ basis functions; this number varied across different datasets and was selected automatically by the function \texttt{auto\textunderscore basis} available from the \texttt{R} package \texttt{FRK} \citep{zammit2021frk}. Here the parameter estimation is done using an expectation-maximization algorithm implemented in the function \texttt{SRE.fit}. Setting the mean process to zero is not allowed in the current version of the \texttt{FRK} package and we chose a constant mean in our implementation; however, this approach is applicable for spatial prediction only and we cannot use it for estimating the Mat\'ern covariance-related parameters. A single core was used for the computation. 






\subsubsection{Gapfill}
Gapfill \citep{gerber2018predicting} is originally a purely algorithmic and nonparametric approach for kriging only. It is specifically attractive due to its lower computational burden, and the prediction is done separately at each spatial location using only a subset of the data. Thus, we can do the missing value imputations in parallel; to predict the response at an unobserved location $\bm s^*$, the Gapfill method defines a spatial neighborhood $\mathcal{N}^b(\bm s^*)$ around $\bm s^*$ and selects a subset $\mathcal{A}=\{Y(\bm s_i): \bm s_i \in \mathcal{N}^b(\bm s^*)\}$
based on which then a prediction $Y(\bm s^*)$ is obtained using sorting algorithms and quantile regression. The selection of subset $\mathcal{A}$ is crucial; it has to be small to gain computationally but, at the same time, it should contain enough observations for good accuracy. The original method does uncertainty quantification using permutation algorithms \citep{gerber2018predicting}. While \texttt{R} package \texttt{gapfill} \citep{gerber2017Gapfill} implements this method, the package allows only a spatiotemporal gridded data structure. However, the KAUST competition datasets are neither spatiotemporal nor gridded.

We implemented a simplified parametric version of this method; around each prediction location $\bm{s}^* = (s_X^*, s_Y^*)'$, we selected a square neighborhood $(s_X^*-\Delta, s_X^* + \Delta) \times (s_Y^*-\Delta, s_Y^* + \Delta)$, where the choice of $\Delta$ is problem specific and we treated it as a tuning parameter. Because the KAUST competition datasets were generated from GPs, we simply fit a Gaussian distribution to all the observations within the neighborhood, instead of using a nonparametric quantile regression technique. The kriging across the prediction locations can be performed independently and we did it in parallel over 20 cores. 


\subsection{Bayesian methods} 
\label{subsec:Bayesian.methods}
The SLME model \eqref{eq:mixedeffects.model} can be rewritten as a hierarchical model as
\begin{align}
\label{eq:BHM}
\bm{Y}\sim \mathcal{N}_N(\mathbf{X}\bm \beta + \bm{W} , \tau^2 \mathbf{I}_N); \quad
\bm{W}  \sim \mathcal{N}_N(\bm 0, \mathbf{C}_{\bm \theta_c}); \quad \bm \beta \sim \mathcal{N}_P(\bm 0, \sigma_{\beta}^2 \mathbf{I}_P); \quad
\bm \theta \sim \pi(\bm\theta),
\end{align}
where $\bm{W}=[\bm{W}(\bm s_1),\ldots,\bm{W}(\bm s_N)]'$, $\mathbf{X}$ is a $(N\times P)$-dimensional design matrix as in \eqref{eq:mixedeffects.model}, $\mathbf{C}_{\bm\theta_c}$ is the $(N\times N)$-dimensional covariance matrix such that $\bm \Sigma_{\bm \theta}  = \mathbf{C}_{\bm \theta_c} + \tau^2 \mathbf{I}_N$ with $ \bm \Sigma_{\bm \theta}$ being the covariance matrix with elements given by \eqref{eq:matern_correlation}, and $\bm \theta_c =(\sigma^2, \phi, \nu)'$ with $\bm\theta=(\bm \theta_c, \tau^2)'$ being the hyperparameters of the model with prior distribution given by $\pi(\bm \theta)$. In the Bayesian paradigm, the inferential problem translates into getting the posterior distribution of the model hyperparameters $\bm \theta$, latent spatial effects $\bm{W}$, and regression coefficients $\bm \beta$. Below we detail the two popular Bayesian inferential methods for model \eqref{eq:mixedeffects.model}. 


\subsubsection{Stochastic partial differential equation (\texttt{SPDE})}
\label{subsec:spde}
The stochastic partial differential equation (SPDE) approach in spatial statistics is based on the link between GPs with Mat{\'e}rn covariance and the Gaussian Markov random fields \citep{lindgren2011explicit}. The implementation in \texttt{R-INLA} is done by projecting the SPDE  onto a basis representation $W(\bm s) = \sum_{k=1}^{K} w_k \psi_k(\bm s)$, where $\psi_k(\cdot)$ are deterministic piecewise polynomial basis function with compact support (to preserve the Markov structure) and $w_k, k=1,\ldots,K$ are the basis weight functions chosen so that the distribution of $W(\bm s)$ approximate the distribution of solution to the SPDE in fixed domain. 

Modeling with \texttt{R-INLA} starts with the triangulation of the spatial domain using \texttt{inla.mesh.2d} function that returns three sparse precision matrices $\bm{C}$, $\bm{G}_1$, and $\bm{G}_2$ along with the sparse projection matrix $\bm{A}$ that projects the mesh nodes (basis locations) to the observed spatial data locations. For a two-dimensional spatial domain and smoothness parameter $\nu=1$ in \eqref{eq:matern_correlation}, an explicit expression of the appropriate sparse precision matrix for the vector of the weight parameters $w_k$'s in terms of Mat\'ern range $\phi$ and marginal variance $\sigma^2$ is given by
$$
\bm{Q} = {{\phi^2}\over{\sigma^2\sqrt{4\pi}}} \left({{1}\over{\phi^4}} \bm{C} + {{2}\over{\phi^2}} \bm{G_1} + \bm{G_2}\right),
$$
and the covariance matrix $\bm{C}_{\theta_c}$ in \eqref{eq:BHM} is well approximated by $ \bm{A} \bm{Q}^{-1} \bm{A} \approx \bm{C}_{\theta_c} $. The sparse structure of $\bm{Q}$ reduces the computational complexity involved in calculating \eqref{eq:multivariate_normal} to $\mathcal{O}(N^{3/2})$.


For the KAUST competition and the real data application in Section \ref{sec:Data-Description}, we followed the step-by-step implementation of the SPDE approach by \cite{lindgren2015bayesian} with the smoothness parameter fixed to $\nu=1$. While performing predictions and estimation, the tuning parameter of main \texttt{R-INLA} functions were set to default values. Finally, we calculated the prediction standard deviation by first extracting it from the \texttt{R-INLA} outputs and then combining it with the estimated nugget $\tau^2$. We used a single core for the computation. 


\subsubsection{Nearest Neighbor Gaussian Process (\texttt{NNGP})}
The nearest neighbor Gaussian process (NNGP) of \cite{datta2016hierarchical, datta2016nonseparable} exploits the full-conditional specification of the joint distribution along with the predictive processes \citep{banerjee2008gaussian} to provide an accurate and faster inference of model \eqref{eq:mixedeffects.model}. The NNGP approach consists of mainly two methods for the inference; namely  \textit{conjugate} and the \textit{response} methods.
Integrating out $\bm{W}$ in \eqref{eq:BHM}, we get
\begin{align}
\label{eq:responsemodel}
\bm{Y}\sim \mathcal{N}_N(\mathbf{X}\bm \beta, \bm \Sigma_{\bm \theta} ), \quad \bm \Sigma_{\bm \theta} = \mathbf{C}_{\bm\theta_c}+\tau^2\mathbf{I}_N.
\end{align}
Since the covariance function of an NNGP can be derived from any parent GP \citep{finley2019efficient}, the covariance matrix $\bm \Sigma_{\bm\theta}$ in \eqref{eq:responsemodel} may be replaced with its NNGP analog $\widetilde{\bm \Sigma}_{\bm \theta}$, which yield the \textit{response} NNGP model $
\bm{Y}\sim \mathcal{N}_N(\mathbf{X}\bm\beta, \widetilde{\bm \Sigma}_{\bm \theta})$; see \cite{datta2016hierarchical, datta2016nonseparable} and \cite{finley2019efficient} for more details. 
On the other hand, the \textit{conjugate} NNGP algorithm is a simplified version of the \textit{response} NNGP model where we fix spatial covariance parameters leading to exact (MCMC free) posterior inference. In particular, if we consider $\bm \Sigma_{\bm\theta}$ as  Mat\'ern covaraince function \eqref{eq:matern_correlation} with marginal variance $\sigma^2$, range $\phi$, smoothness $\nu$, and nugget $\tau^2$, with $\alpha=\tau^2/\sigma^2$, then the \textit{conjugate} NNGP model fixes $\phi$, $\alpha$, and $\nu$ and generates the NNGP covariance matrix approximation $\tilde{\bm{M}}$.  Hence, we obtain the following marginal model $\bm{Y}\sim \mathcal{N}_N(\mathbf{X}\bm \beta, \sigma^2\tilde{\bm{M}})$,
where $\tilde{\bm{M}}\equiv \tilde{\bm{M}}(\alpha,\phi,\nu)$ is a known covariance matrix. We may use Gaussian and inverse-gamma conjugate priors for $\bm \beta$ and $\sigma^2$, respectively, which simplifies the calculation as posterior moments and other summary statistics can be obtained easily and exactly (MCMC-free inference). Also, in this situation, the posterior distribution of $Y(\bm s^*)$ at a new location $\bm{s}^*$ has closed-form expressions hence facilitating fast computation.

We used the \textit{conjugate} NNGP method by exploiting the function \texttt{spConjNNGP} from \texttt{R} package \texttt{spNNGP} \citep{finley2022spnngp}. We chose the neighborhood size of $50$ for each of the datasets where we discretized the spatial covariance parameters $\nu,\phi,$ and $\alpha$ in their respective domains. More specifically, we chose four equally-spaced sequences from the intervals $(0.01,1)$ for $\alpha$, $(0.1,10)$ for $\phi$, and $(0.1, 5)$ for $\nu$, resulting in total $4^3=64$ combinations of parameters. We used 5-fold cross-validation to obtain the optimal estimates. We generated $200$ posterior samples from each of the parameters and obtained the posterior estimates of all the model parameters $\bm \beta$, $\sigma^2, \alpha, \phi$, and $\nu$. We used an intercept, latitude, and longitude as the components of spatial trend in \eqref{eq:mixedeffects.model}; as of our knowledge, the mean function cannot be set to zero. After the parameter estimates are obtained, we used the same model to do the prediction for all the datasets using the function \texttt{predict} from \texttt{spNNGP}, where the estimates of the prediction standard deviations were also reported. We used a single core for the computation; the parallelization requires \texttt{OpenMP} \citep{dagum1998openmp}.


\subsection{Deep learning models}
\label{subsec:deepL.methods}
{\color{black}{Geostatistics has witnessed the emergence of several spatial deep learning models designed to capture intricate spatial dependencies and patterns in spatial data; including Deep Gaussian Processes \citep[DeepGPs,][]{pmlr-v31-damianou13a}--an extension of GPs to multiple layers, Spatial Convolution Neural Networks \citep{shi2015convolutional}--a variant of convolution neural network that is flexible in capturing spatial dependence, Graph Neural Networks \citep{wu2020comprehensive}--a neural network designed to handle non-lattice spatial data, and deep kriging \citep[DeepKriging,][]{chen2020deepkriging}--an extension of the concept of classical spatial kriging to deep layers. For a comprehensive review of such approaches, see \cite{wikle2023statistical}.
In this section, we mainly focus on two well-known deep learning models that use GPs to model complex relationships in data and provide uncertainty estimates in predictions.}} 
\subsubsection{\bf Deep Gaussian Process (\texttt{DeepGP})}
Deep Gaussian Processes \citep[DeepGPs,][]{pmlr-v31-damianou13a} are a class of deep belief networks that leverage GP mappings. This framework represents data as the output of a multivariate GP, while the inputs to this process are governed by another GP, enabling a hierarchical modeling strategy. \textcolor{black}{More explicitly, an $L$-layer DeepGP can be represented as
\begin{align}
\label{eq:DeepGP-general}
   \bm{f}^{(l)}=\bm{f}^{(l-1)} + \bm{u}^{(l)}, \quad l=0,\ldots,L,
\end{align}
where $\bm{f}^{(l)} = [f^{(l)}_1, f^{(l)}_2, \ldots, f^{(l)}_N]$ is the vector of latent function values at layer $l$, $\bm{f}^{(0)} = \mathbf{X}$ is the input data (observed covariates and the spatial basis functions), and $\bm{u}^{(l)} = [u^{(l)}_1, u^{(l)}_2, \ldots, u^{(l)}_N]$ is a vector of realizations from a GP with zero mean and a covariance kernel $K^{(l)}(\mathbf{x}, \mathbf{x}')$, e.g., a Mat\'ern kernel \eqref{eq:matern_correlation}, and it is independent of $\bm{f}^{(l-1)}$. The latent function $\bm{f}^{(L)}$ is the last product of the DeepGP at layer $L$, and it is utilized to model the observed data $\bm{y}$ via a likelihood function.} By incorporating multiple layers of this GP regression, \cite{pmlr-v31-damianou13a} introduced a novel non-Gaussian modeling framework; see also \cite{wilson2016deep, salimbeni2017doubly} for further advancement of this topic. To draw inference for this model, approximate variational marginalization is employed, yielding a strict lower bound on the marginal likelihood, which serves as a valuable criterion for model selection, helping identify the optimal number of layers and nodes per layer.

A single-layer DeepGP model \eqref{eq:DeepGP-general} within this framework is equivalent to a standard GP or the latent variable model, \textcolor{black}{ and a two-layer model may be interpreted as the SLME model \eqref{eq:mixedeffects.model}, when appropriately formulated. More explicitly, \eqref{eq:mixedeffects.model} can be obtained from a two-layered DeepGP \eqref{eq:DeepGP-general} where the first layer output is assumed to be GP and the second layer is assumed as a final output. We can formulate it as $\bm{f}^{(1)}= \bm{u}^{(1)}$ and $\bm{f}^{(2)}=\bm{f}^{(1)} + \bm{u}^{(2)}$.
We may interpret $\bm{f}^{(1)}$ and $\bm{f}^{(2)}$ as the spatially correlated random effects $W(\bm s)$ and responses $Y(\bm s)$ in \eqref{eq:mixedeffects.model}, respectively, hence linking the two approaches explicitly. 
 Therefore, in general, DeepGPs are more flexible and can model nonlinear relationships between variables due to the composition of multiple GP layers. They can automatically learn hierarchical representations of data, allowing them to capture complex spatial structures and nonlinear spatial dependencies.}



We utilized DeepGPs (available as a package \texttt{sklearn} in \texttt{python}) to predict processes at unobserved locations. By embedding the locations as basis functions to approximate $W(\bm{s})$, as defined in \eqref{eq:basis}, we harnessed the power of DeepGPs for accurate and efficient spatial predictions. The computation is highly parallelizable and we used 40 cores here.

\subsubsection{\bf DeepKriging} \label{deepkriging}
\textcolor{black}{Just as DeepGPs extend GPs to several layers within the context of deep learning, DeepKriging also extends the concept of (universal) kriging to multiple levels. The (universal) kriging prediction for the SLME model \eqref{eq:mixedeffects.model} is $\hat{Y}(\bm s^*) = \mathbf{X}(\bm s^*)^{'} \hat{\bm\beta} + \bm{c}(\bm s^*)^{'}\bm\Sigma_{\theta}^{-1} (\bm y - \mathbf{X}\hat{\bm\beta}),$
where $\bm{c}(\bm s^*) = \text{Cov}(\bm{Y}, Y(\bm s^*))$, $\mathbf{X} = (\mathbf{X}(\bm s_1), \ldots, \mathbf{X}(\bm s_1))^{'} $ is $N\times P$ design matrix, and $\hat{\bm\beta}= (\mathbf{X}^{'} \bm\Sigma_{\theta}^{-1} \mathbf{X}^{'})^{-1} \mathbf{X}^{'} \bm\Sigma_{\theta}^{-1} \bm{y}$. The spatial dependence is incorporated in $\hat{Y}(\bm s^*)$ through a linear function of the covariance vector $\bm{c}(\bm s^*)$, where the coefficient $\bm \Sigma^{-1}(\bm{y} - \mathbf{X}\hat{\bm\beta})$ remains unknown, and often costly to calculate. 
This inspires us to employ a set of established nonlinear functions as embeddings of the spatial locations $\bm{s}$ in the features, allowing us to effectively characterize the spatial process within the neural network. Utilizing the Karhunen-Lo\'eve expansion of a random field, the spatial process $W(\bm s)$ in \eqref{eq:mixedeffects.model} can be represented as  
    \begin{equation}
         W(\bm s) = \sum_{k=1}^{\infty} w_{k}~\psi_{k}(\bm{s}),
         \label{eq:basis}
    \end{equation}
where $w_{k}$'s are independent random variables and $\psi_{k}(s)$'s are pairwise orthonormal basis functions. The process \eqref{eq:basis} can be approximated as $W(\bm s) \approx \overline{W}(\bm s) = \sum_{k=1}^{K} w_{k}~\psi_{k}(\bm{s})$, where the weights $w_{k}$ can be estimated by minimizing the mean square error.} 

This idea is formulated in DeepKriging, first proposed by \cite{chen2020deepkriging}, which reshapes the spatial interpolation problem as regression and introduces a spatially dependent neural network structure for spatial prediction, uses Wendland radial basis functions to embed the spatial locations into a vector of weights and passes them to the deep neural networks as inputs allowing them to capture the spatial dependence of the process. 
In DeepKriging, the prediction intervals are obtained using bootstrap samples \citep{srivastav2007simplified,boucher2009tools}. For the univariate scenario, the prediction interval is given by
    $
    \hat{Y}_{mean}(\bm{s}^*) \pm t_{(1-\alpha/2),df} \sqrt{\hat{\sigma}_{model}^2(\bm{s}^*) + \hat{\sigma}_{\varepsilon}^2(\bm{s}^*) } , 
$
where $\hat{Y}_{mean}(\bm{s}^*)$ is the mean of the predictions from ensembles, $\hat{\sigma}_{model}^2(\bm{s}^*)$ is the estimated model variance, and $\hat{\sigma}_{\varepsilon}^2(\bm{s}^*)$ is the estimated variance of the random noise $\varepsilon$.


We implemented this framework in \texttt{python}, using a total of $K = 1830$ basis functions with a similar architecture as in \cite{nag2023bivariate}. Similar to DeepGP, the computation is highly parallelizable, and we utilized 40 cores for the implementation. The code for our implementation is available on GitHub \href{https://github.com/pratik187/CaseStudy_spatial}{here}.

\section{Comparative study and results}
\label{sec:comparison_results}
\subsection{Application to KAUST competition datasets}
 In this study, we extensively applied the methods discussed in Section \ref{sec:statistical_models} to analyze the Sub-competition 2a datasets (see Section \ref{sec:competition}). To evaluate the performance of each method, we conducted a comparative 3-fold cross-validation study, and the results are presented in Table \ref{table:simulated_data}. It is evident that most of the statistical methods such as BCL, GpGp, GpGp0, SPDE, and NNGP outperformed the deep learning (semiparametric) and Gapfill (algorithmic) approaches, and provided consistent results across all five datasets irrespective of different spatial patterns and dependence structures. Two possible reasons are: (1) These models have underlying Gaussian assumptions that match the true data-generating distribution, and (2) The nonparametric or semiparametric approaches do not use the information that the covariance structure used for simulating the competition datasets is isotropic Mat\'ern in \eqref{eq:matern_correlation}. \textcolor{black}{FRK (semiparametric) through basis functions provides a low-rank approximation of the covariance function. It is a linear model, and the results showed its underperformance in spatial prediction. A nonlinear version of FRK, i,e., DeepKriging, performed better here. These differences are more evident for Datasets 3 and 5, which pose weaker dependence (small range) than other datasets. In particular, among proper statistical models, BCL, GpGp, GpGp0, and NNGP provided very similar results across five different datasets.}
 \begin{table}[]
\setlength\extrarowheight{-5pt}
\addtolength{\tabcolsep}{-2pt}
\caption{Comparison of different models (using a 3-fold cross-validation) based on multiple scores for the five datasets of Sub-competition 2a. The horizontal blocks of the table represent the results for datasets 1--5. Here MSPE, MAPE, PICP, MPIW, and Time (s) stand for mean squared prediction error, mean absolute prediction error, 95$\%$ prediction interval coverage probability, mean 95$\%$ prediction interval width, and computation time in seconds, respectively. Here, a single core was used for FRK, GpGp, GpGp0, NNGP, and SPDE, 20 cores were used for BCL and Gapfill, and 40 cores were used for DeepKriging and DeepGP.}
\label{table:simulated_data}
\begin{tabular}{cccccccccc}
\hline \vspace{-5mm}\\ 
Models   & BCL   & FRK   & Gapfill & GpGp  & GpGp0 & SPDE  & NNGP  & DeepKriging & DeepGP \\
\hline
MSPE     & 0.274 & 0.277 & 0.282   & 0.274 & 0.274 & 0.274 & 0.278 & 0.318       & 0.317  \\
MAPE      & 0.417 & 0.420 & 0.424   & 0.417 & 0.417 & 0.417 & 0.420 & 0.449       & 0.448  \\
PICP     & 0.951 & 0.951 & 0.964   & 0.953 & 0.953 & 0.951 & 0.951 & 0.939       & 0.961  \\
MPIW     & 2.049 & 2.062 & 2.258   & 2.075 & 2.074 & 2.049 & 2.064 & 2.144       & 2.310  \\
Time (s) & 462   & 60    & 8       & 547   & 595   & 186   & 14654 & 1187        & 4379  \\
\hline \vspace{-5mm}\\ 
MSPE     & 0.274 & 0.306 & 0.347   & 0.275 & 0.275 & 0.276 & 0.279 & 0.313       & 0.297  \\
MAPE      & 0.417 & 0.441 & 0.470   & 0.418 & 0.418 & 0.419 & 0.421 & 0.441       & 0.421  \\
PICP     & 0.951 & 0.951 & 0.983   & 0.953 & 0.952 & 0.951 & 0.952 & 0.942       & 0.961  \\
MPIW     & 2.053 & 2.169 & 3.246   & 2.076 & 2.076 & 2.058 & 2.076 & 2.159       & 2.330  \\
Time (s) & 393   & 62    & 8       & 510   & 478   & 138   & 14533 & 1259        & 4112 \\
\hline \vspace{-5mm}\\ 
MSPE     & 0.297 & 1.055 & 0.984   & 0.297 & 0.297 & 0.419 & 0.298 & 0.380       & 0.829  \\
MAPE      & 0.434 & 0.818 & 0.788   & 0.434 & 0.434 & 0.515 & 0.436 & 0.490       & 0.721  \\
PICP     & 0.951 & 0.949 & 0.974   & 0.952 & 0.952 & 0.949 & 0.954 & 0.951       & 0.963  \\
MPIW     & 2.138 & 4.029 & 4.391   & 2.147 & 2.147 & 2.538 & 2.177 & 2.418       & 3.830  \\
Time (s) & 549   & 61    & 8       & 514   & 521   & 160   & 14966 & 1223        & 4719  \\
\hline \vspace{-5mm}\\ 
MSPE     & 0.275 & 0.307 & 0.346   & 0.275 & 0.275 & 0.277 & 0.279 & 0.319       & 0.296  \\
MAPE      & 0.418 & 0.441 & 0.468   & 0.418 & 0.418 & 0.419 & 0.421 & 0.447       & 0.433  \\
PICP     & 0.951 & 0.950 & 0.967   & 0.953 & 0.953 & 0.951 & 0.951 & 0.951       & 0.955  \\
MPIW     & 2.055 & 2.169 & 2.556   & 2.077 & 2.077 & 2.060 & 2.071 & 2.141       & 2.200  \\
Time (s) & 541   & 61    & 8       & 528   & 494   & 152   & 14666 & 1355        & 4139  \\
\hline \vspace{-5mm}\\ 
MSPE     & 0.302 & 1.185 & 1.004   & 0.302 & 0.302 & 0.455 & 0.304 & 0.409       & 0.956  \\
MAPE      & 0.438 & 0.863 & 0.797   & 0.438 & 0.438 & 0.537 & 0.440 & 0.509       & 0.779  \\
PICP     & 0.951 & 0.947 & 0.976   & 0.951 & 0.952 & 0.951 & 0.955 & 0.962       & 0.964  \\
MPIW     & 2.154 & 4.259 & 4.576   & 2.162 & 2.162 & 2.649 & 2.206  & 2.493       & 4.088  \\
Time (s) & 427   & 61    & 8       & 565   & 585   &  167   & 15099 & 1457        & 4113 \\
\hline
\end{tabular}
\end{table}
Moreover, the SPDE approach demonstrates highly competitive results when applied to datasets with strong spatial dependence (Datasets 1, 2, and 4). However, it tends to underperform when dealing with datasets exhibiting relatively weaker spatial dependence (Datasets 3 and 5). \textcolor{black}{This discrepancy in performance could potentially be attributed to the rigid fixation of the smoothness parameter in the SPDE approach (see Section \ref{subsec:spde}) at a constant value of one, which may not accurately capture the true underlying smoothness characteristics of the data. Adjusting the smoothness parameter to better align with the specific spatial characteristics of each dataset might lead to improved performance across all datasets.} The coverage probabilities of the $100(1-\alpha)\%$ prediction intervals for all these approaches, with $\alpha=0.05$, are close to 95$\%$ with narrower interval width for the proper statistical models compared to  FRK, deep learning models, and Gapfill. Notably, deep learning models perform well compared to FRK and Gapfill. Among the two deep learning models, DeepKriging provides more stable results than DeepGP; the latter performs the worst for datasets with weak dependence (Datasets 3 and 5). Thus, based on the performance and stability of these approaches across five different datasets, we may reorder them as follows
 $$
 \text{BCL, GpGp, GpGp0, NNGP} > \text{SPDE} > \text{DeepKriging} > \text{DeepGP} > \text{FRK} > \text{Gapfill}. 
$$
Overall, BCL, GpGp, GpGp0, and NNGP are the most stable and best-performing approaches, irrespective of different spatial patterns and dependence structures across datasets. Hence, they are trustworthy for the prediction across spatial locations with varying dependence strength and patterns under the KAUST competition settings. The computation time is reasonable for all the approaches, though NNGP is relatively costly and the cost may be reduced by changing the discretization of the parameter space and the neighborhood size.

Upon careful examination of the outcomes, we chose zero-mean GpGp, i.e., GpGp0 as the best model, though it performs very similarly to GpGp in cross-validation. For support in favoring GpGp0 as a better option than GpGp, in Table \ref{table:Summary-Mean}, we report the summary of the estimated spatial trend (mean) component using GpGp for the sub-challenge 1a and 1b detailed in Section \ref{sec:competition}, where the true mean is zero. From Table \ref{table:Summary-Mean}, we observe that for Datasets 1, 2, and 4, the estimated means significantly differ from zero, though for Datasets 3 and 5, the estimates are not significant. This is possibly due to the weak spatial dependence in Datasets 3 and 5 that provides favorable conditions to apply the central limit theorem (CLT) under a dependent sequence of random variables, hence driving the estimates close to the population mean (zero). Nevertheless, the nonzero estimates of the mean component, in general, motivated us to fix the mean to zero beforehand, which is not allowed in the current version of the \texttt{GpGp} package (0.4.0). Based on these observations, we decided to focus on employing GpGp0 to obtain the final set of results for the competition. The successful implementation of GpGp0, using the necessary \texttt{R} and \texttt{C++} elements from the package \texttt{GpGp}, played a crucial role in winning two sub-competitions 1b and 2a (see details \href{https://cemse.kaust.edu.sa/stsds/news/2023-kaust-competition-spatial-statistics-large-datasets}{here}). In the subsequent subsection, we explain the detailed implementation of GpGp0, exploring its key components and specific parameter choices.




\subsection{Details about the final model: zero-mean GpGp (\texttt{GpGp0})
}
\label{subsec:GPGP}

\begin{table}[h]
\centering
\setlength\extrarowheight{-5pt}
\addtolength{\tabcolsep}{-2pt}
\caption{Estimates of the constant spatial mean parameter $\mu$ using the function \texttt{fit\textunderscore model} from the \texttt{R} package \texttt{GpGp} for the five training datasets under the Sub-competitions 1a and 1b. The corresponding standard errors, $Z$-scores (assuming normality of the estimator), and $p$-values are also reported. The original value of $\mu$ is zero.}
\label{table:Summary-Mean}
\begin{tabular}{ccccccc}
\hline
Sub-competition     & Metric     & Dataset 1 & Dataset 2 & Dataset 3 & Dataset 4 & Dataset 5 \\
\hline
\multirow{4}{*}{1a} & $\hat{\mu}$    & -0.6079   & -0.2785   & -0.0536   & -0.5502   & 0.0124    \\
                    & $\textrm{SE}(\hat{\mu})$ & 0.2086    & 0.2573    & 0.0889    & 0.1641    & 0.102     \\
                    & $Z$-score    & -2.9145   & -1.0823   & -0.6023   & -3.3522   & 0.1212    \\
                    & $p$-value    & 0.0036    & 0.2791    & 0.5469    & 0.0008    & 0.9036    \\
                    \hline
\multirow{4}{*}{1b} & $\hat{\mu}$    & -1.4541   & -0.8633   & 0.0897    & -0.9741   & -0.075    \\
                    & $\textrm{SE}(\hat{\mu})$ & 0.157     & 0.0979    & 0.0597    & 0.1184    & 0.0521    \\
                    & $Z$-score    & -9.2621   & -8.8209   & 1.5036    & -8.2265   & -1.4398   \\
                    & $p$-value    & $2.0\times 10^{-20}$         & $1.1\times 10^{-18}$         & 0.1327    & $1.9\times 10^{-16}$         & 0.1499   \\
                    \hline
\end{tabular}
\end{table}




In the competition, our team \textit{DesiBoys} emerged victorious in Sub-competitions 1b and 2a. Sub-competition 1b focused on constructing confidence intervals for Mat\'ern covariance parameter estimates. For this purpose, we employed the zero-mean Vecchia approximation methodology to estimate the parameters. Initially, we obtained parameter estimates for each entire dataset. While \texttt{GpGp0} (via \texttt{GpGp}) returns the Fisher information matrix of the Mat\'ern covariance parameters, a Gaussian approximation to obtain 95\% confidence intervals puts a significantly large weight to negative values of the Mat\'ern parameters for all datasets, which is practically not possible. Hence, we consider a (approximate) bootstrap approach despite its huge computational burden. The initial estimates were used to conduct 1000 simulations on a subsample of 10,000 points from the given locations. Due to the nonhomogeneous spatial point patterns as seen in Figure \ref{fig:train_test_maps} (spatial point patterns in the training datasets in Sub-competitions 1a/2a and 1b/2b are similar), we assigned unequal weights to each location to obtain the subsamples, aiming to de-cluster the dataset. Specifically, the weight assigned to a particular point was chosen as the inverse of the total number of points in its small neighborhood. Subsequently, the weights were normalized to ensure they added up to one. In the function \texttt{fit\textunderscore model\textunderscore meanzero}, we employed the \texttt{GpGp} configurations with 1000 simulations for each dataset, using \texttt{m\textunderscore seq}= c(10, 30, 60) defining a sequence of values for the number of neighbors. Following the simulations, we fitted a skew-normal distribution for each parameter estimate obtained from the 1000 simulations; this is done specifically to obtain smoother estimates of the sampling distributions. To accomplish this, we utilized the \texttt{selm} function from the \texttt{R} package \texttt{sn}. The 95\% confidence intervals were obtained from the quantiles (0.025$^{th}$ and 0.975$^{th}$) using the \texttt{qsn} function from the package \texttt{sn}. 

For Dataset 2a, we employed the Vecchia approximation (with zero means) to estimate the parameters initially and then empirically calculated pointwise 95\% prediction intervals. The kriging predictions were obtained at the prediction locations, and since the marginal distributions of the predictive samples are Gaussian for a Gaussian process, knowing the prediction variances (in addition to the prediction means) sufficed to derive the 95\% pointwise prediction intervals. To achieve this, we utilized the Vecchia approximation to simulate predictive samples and calculated sample prediction variances. We followed a strategy similar to Competition 1b for the estimation process. The tuning parameters were set to the default choices of the \texttt{fit\_model} function in the \texttt{GpGp} package, with \texttt{m\_seq }= c(10, 30, 60). To obtain the kriging predictions, we developed a function \texttt{predictions\_meanzero} that works similarly to the function \texttt{predictions} from the \texttt{GpGp} package along with allowing a zero mean, and for conditional simulation, we employed the \texttt{cond\_sim} function from the same package. To accommodate the larger number of observation locations in our competition, we modified the \texttt{predictions\_meanzero} functions to allow reordering (the original function \texttt{predictions} allows reordering only when the number of locations is less than 60,000), using 200 nearest neighbors, and generating 1000 samples for the numerical approximation of the prediction variance using the function \texttt{cond\_sim}. We have made the codes for this implementation available via our GitHub repository \href{https://github.com/pratik187/CaseStudy_spatial.git}{here}. 


\section{Application to the total precipitable water data}
\label{sec:Data-Description}



\subsection{Description of the  dataset}
\label{subse:precip_data}
\begin{figure}[h!]
     \centering
     \includegraphics[width=0.325\textwidth]{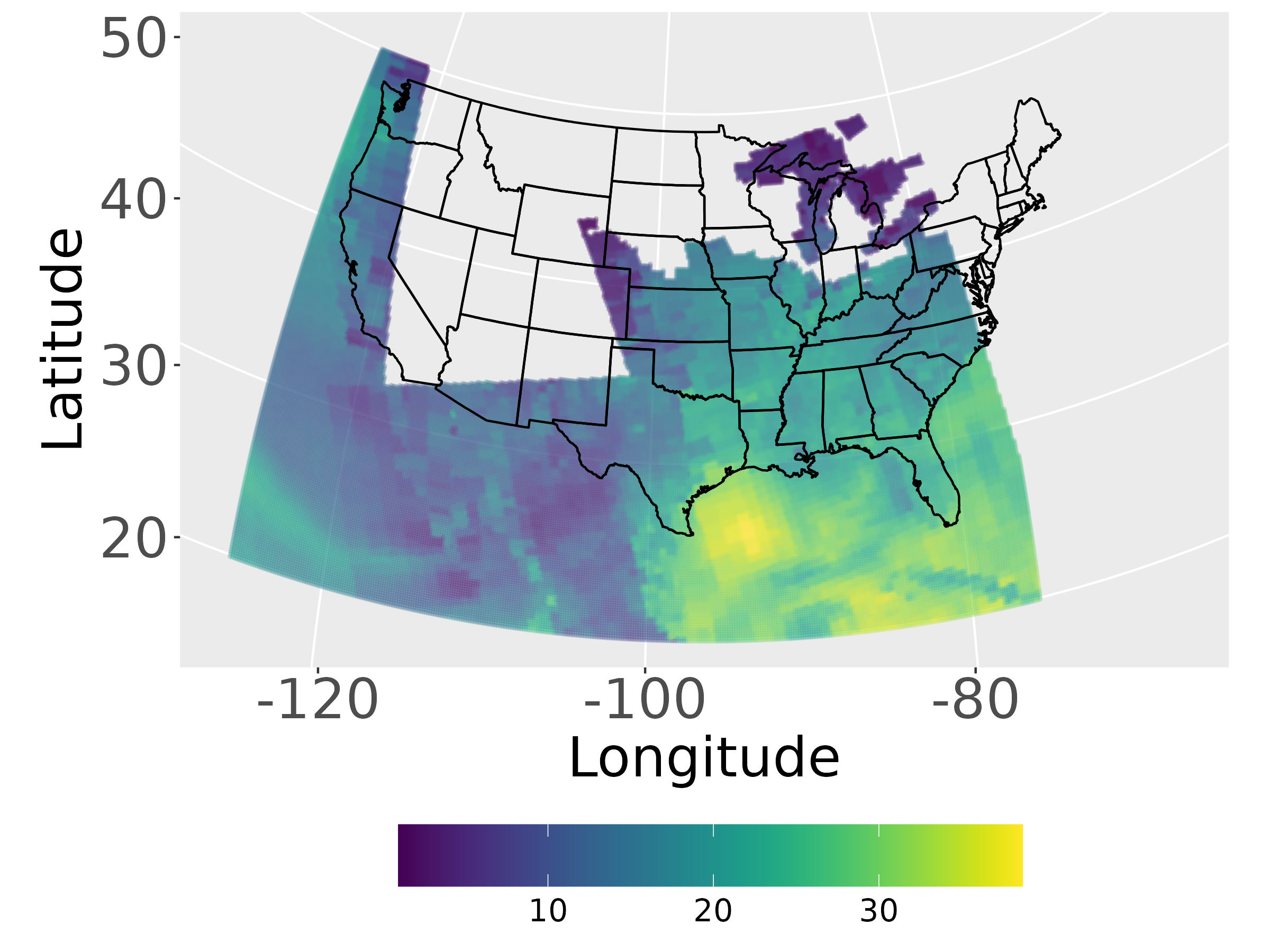}
     \includegraphics[width=0.325\textwidth]{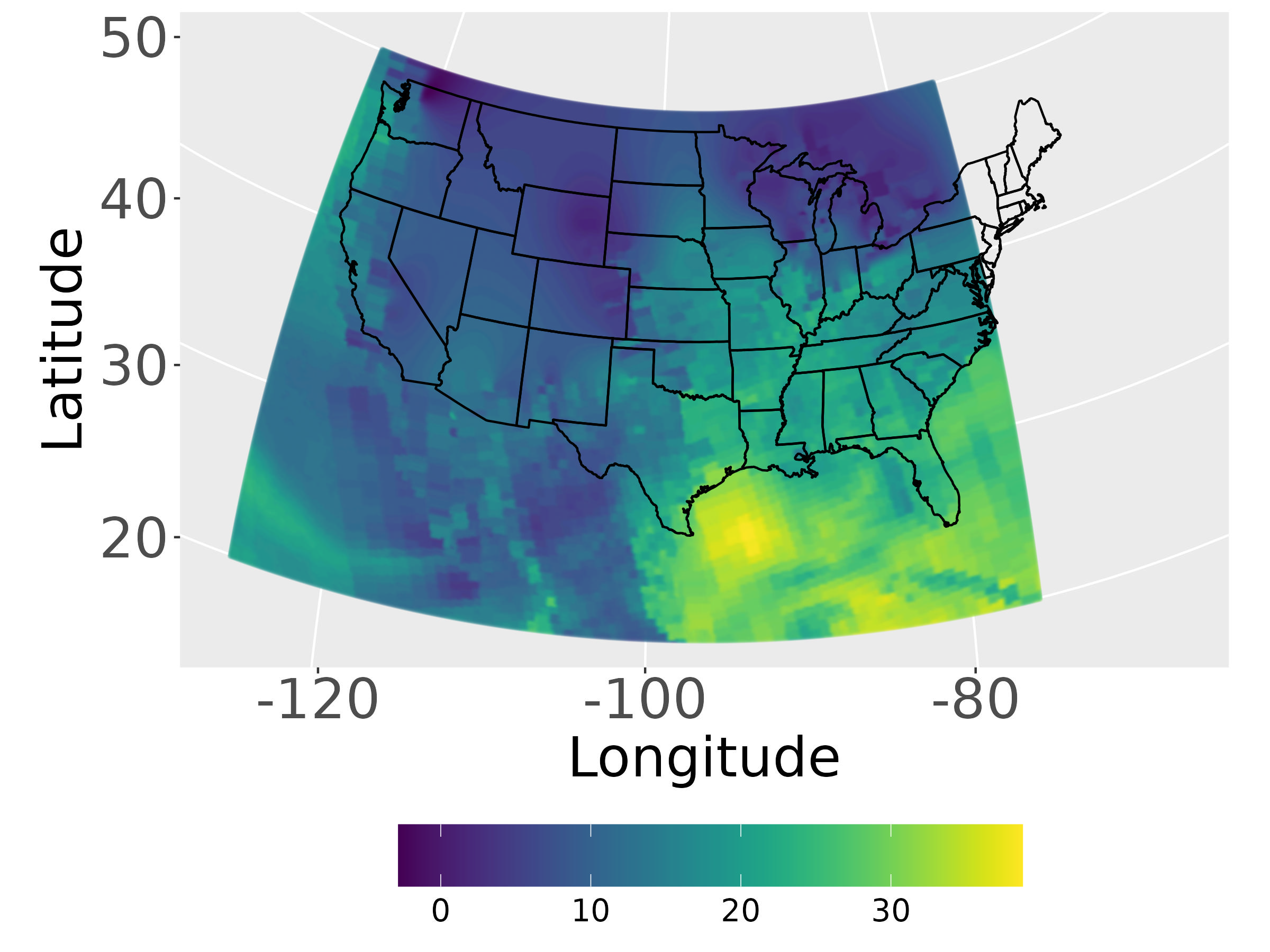}
     \includegraphics[width=0.325\textwidth]{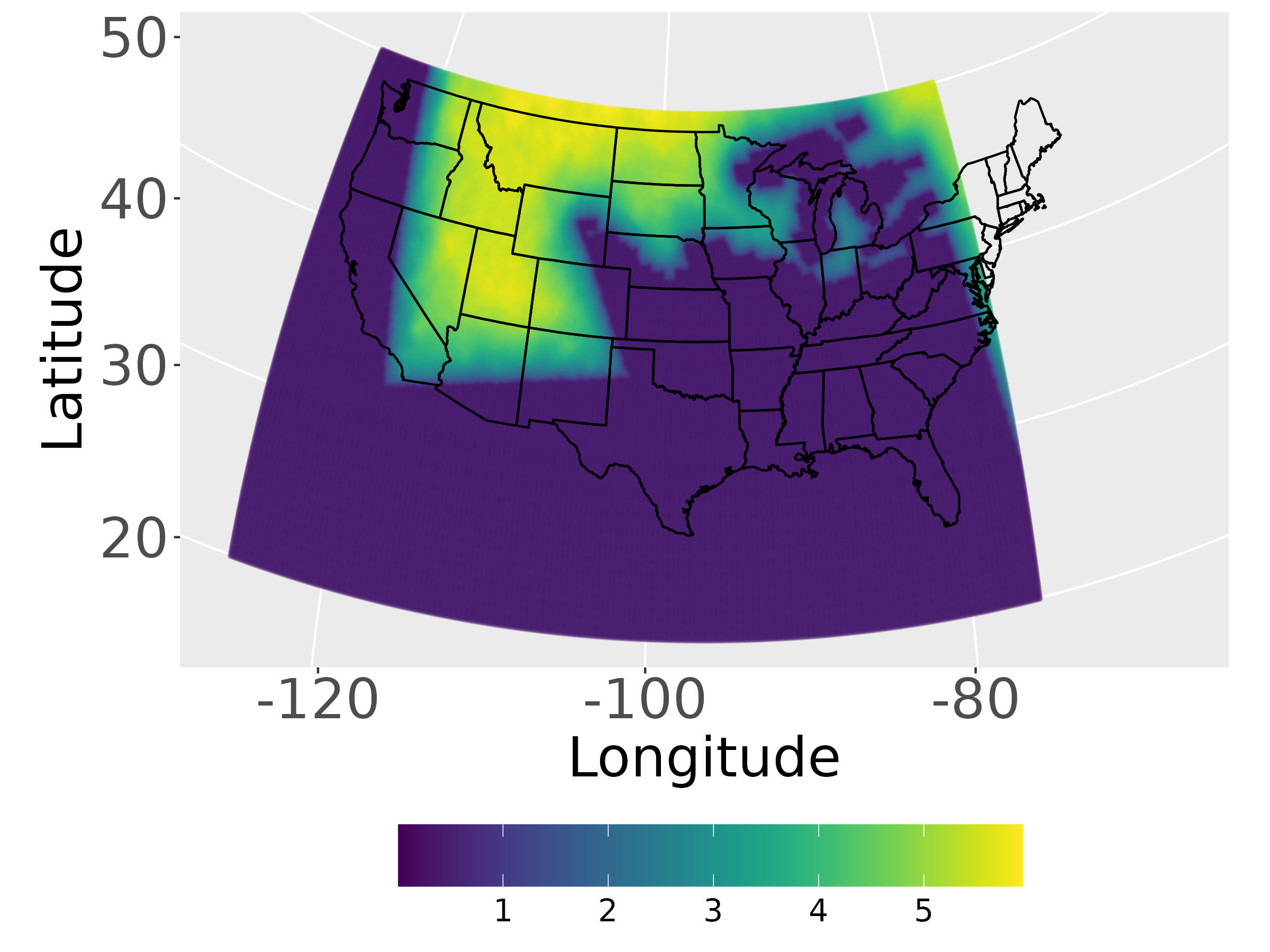}
    \caption{Left panel: total precipitable water (TPW) measurements at 59,754 cells on a $0.25^\circ \times 0.25^\circ$ grid. Middle and right panels: The predictive means and standard deviations on a $0.00025^\circ \times 0.00025^\circ$ grid, obtained using GpGp0. The color scales represent the measurements and predictions in units of mm.} 
    \label{fig:precip_real}
\end{figure}

In this section, we extended the application of the aforementioned methodologies to a substantially large real spatial dataset comprising $n = 59,754$ measurements of total precipitable water (TPW). These measurements were obtained by the Microwave Integrated Retrieval System (MIRS) satellites during a specific time window, from 2 a.m. to 3 a.m. UTC on February 1, 2011, over a region encompassing the United States. We obtained the dataset from \cite{katzfuss2017multi}. The left panel in Figure \ref{fig:precip_real} visually represents the noisy nature of the data and the presence of significant spatial gaps. The current operational practice involves utilizing an ad hoc gap-filled product derived from the data. This product then serves multiple purposes, such as water vapor movement tracking and identifying conditions that may lead to heavy precipitation. The need for a more reliable and robust approach to predict TPW values stems from the noisy and incomplete nature of the original dataset. 

Before we implement some of the approaches described in Section \ref{sec:statistical_models} that fix the mean of the process to be zero (BCL, GpGp0), we first fit a multiple linear regression model with an intercept, longitude, and latitude, and we remove the fitted trend component. Then, we fit the spatial statistical methodologies to the residuals. Further, while reporting the final results, we add that trend component back to the predictions and the predicted intervals.


\begin{table}[h]
\setlength\extrarowheight{-5pt}
\addtolength{\tabcolsep}{-2pt}
\caption{Comparison of different models (using 3-fold cross-validation) based on multiple scores for the real dataset described in Section \ref{sec:Data-Description}. The notations and information on the number of cores used for computation are the same as in Table \ref{table:simulated_data}.}
\label{tab:data-appl}
\begin{tabular}{cccccccccc}
\hline \vspace{-5mm}\\ 
Models   & BCL   & FRK   & Gapfill & GpGp  & GpGp0 & SPDE  & NNGP  & DeepKriging & DeepGP \\
\hline
MSPE     & 0.439 & 2.921 & 1.379   & 0.438 & 0.438 & 0.620 & 0.451 & 0.977       & 0.893  \\
MAE      & 0.366 & 1.245 & 0.769   & 0.366 & 0.366 & 0.475 & 0.376 & 0.637       & 0.773  \\
PICP     & 0.947 & 0.942 & 0.995   & 0.946 & 0.946 & 0.943 & 0.947 & 0.944       & 0.956  \\
MPIW     & 2.651 & 6.703 & 5.079   & 2.635 & 2.634 & 3.133 & 2.691 & 3.420       & 2.977  \\
Time (s) & 336   & 46    & 4       & 299   & 281   & 2783  & 9464  & 1101        & 3973  \\
\hline
\end{tabular}
\end{table}


\subsection{Results}
Table \ref{tab:data-appl} lists the summaries of all results based on fitting the methods described in Section \ref{sec:statistical_models} to the TPW dataset. Likewise, in Section \ref{sec:comparison_results}, we performed 3-fold cross-validations to compare these approaches, and we noticed a very similar comparative performance as in Section \ref{sec:comparison_results}. More explicitly, BCL, GpGp, GpGp0, and NNGP appear to be the best models in terms of lower mean squared prediction error, lower mean absolute error, good coverage probability, and narrower intervals. The SPDE model is the next best model, followed by machine learning models. FRK provided the worst cross-validation score, while Gapfill underperformed equally. The \texttt{SRE.fit} function from the \texttt{FRK} failed to estimate the model parameters by fitting a variogram (shown as a warning); however, the true values of these parameters are not known. Interestingly, the coverage probabilities of the $100(1-\alpha)\%$ prediction intervals for all these approaches, with $\alpha=0.05$, were close to $95\%$, except for Gapfill, which provided almost $100\%$ coverage. After a careful comparison, we chose GpGp0 as the final model that performs close to GpGp but is very slightly better in decimal digits. It is also noteworthy to mention that the GpGp0 is employed for the spatial residuals after removing a spatial trend, as mentioned in Section \ref{subse:precip_data}. 

The middle and right panels in Figure \ref{fig:precip_real} shows the spatial predictive means and standard deviations obtained using the best model GpGp0. It is evident that the estimated mean mimics the true observed TPW in the first row while predicting well at unobserved big white patches. As expected, the prediction standard deviations were larger near the unobserved locations, with smaller values near the observed locations. 

\section{Discussions and conclusions}
\label{sec:conclusions}

This paper provides an extensive overview of statistical and machine learning methods for analyzing large spatial datasets. It proposes a standardized task framework for comparing different analytical approaches, focusing on assessing predictive performance and quantifying prediction uncertainty. Some methods consistently showed better performance on both simulated and real datasets, but their efficacy may vary based on dataset size, measurement-error variance, and missing data. The authors have made the reproducible code publicly available to support implementation in other data analytics projects.



Motivated by the KAUST Large Spatial Data Competition 2023, our team \textit{DesiBoys} thoroughly investigated various statistical models, deep learning techniques, and algorithmic approaches to tackle the inferential challenges posed by large spatial datasets. We considered zero-mean constraints in the data and developed additional \texttt{R} functions to enhance performance. The results showed that proper statistical models consistently outperformed other approaches on all five simulated competition datasets and also the total precipitable water (TPW) dataset. Their performance remained stable and accurate across different competition datasets, despite variations in spatial patterns and dependence strength. On the other hand, deep learning models (DeepKriging and DeepGP), algorithmic methods (Gapfill), and Fixed Rank Kriging (FRK) demonstrated limitations in accurately capturing true spatial patterns, especially when the dependence strength weakened. This study emphasizes the strengths of proper statistical models and the importance of selecting the appropriate method based on dataset characteristics and underlying spatial dependencies.


While our primary focus in this study is assessing predictive performance, future research may extend the analysis to include a comparison of parameter estimation and inference across different approaches. Additionally, an intriguing avenue for investigation involves examining the effects of preferential sampling on the inference for these datasets. Preferential sampling has become a growing concern in spatial and ecological statistics due to its potential to introduce biases stemming from a causal connection between the locations of sample data and the response of interest \citep{pfeffermann2007small}. Therefore, understanding and addressing preferential sampling effects are essential for robust and reliable spatial data analysis and modeling.
One further research avenue would be applying and comparing these models in the spatiotemporal case; there are several flexible spatial models in the literature like the spatial deformation approach \citep{sampson1992nonparametric} and nonparametric Bayesian spatial modeling \citep{gelfand2005bayesian} that only work when temporal replications of the spatial fields are available, and the flexibility of such approaches can be tested as well.

\section*{Acknowledgement}
The first three authors (Hazra, Nag, Yadav) participated in the KAUST 2023 data competition with team name \textit{DesiBoys} and contributed equally to this work by implementing some of the methods and writing the paper. The last author (Sun) oversaw the whole project for the manuscript preparation. 
We would like to thank the KAUST Data Science and Statistics team for organizing this very interesting data competition.


\baselineskip 14pt
\bibliographystyle{CUP}
\bibliography{ref}

\baselineskip 10pt

\end{document}